\documentclass{ws-procs9x6}
\def\VERSION{2.6.X}

\begin{document}

\title{micrOMEGAs : a code for the calculation 
of Dark Matter properties  in  generic models of particle interaction.}

\author{G. B\'elanger and F. Boudjema  }
\address{LAPTH,University de Savoie,CNRS,B.P.110, \\
 F-74941 Annecy-le-Vieux Cedex, France\\
E-mail: belanger@lapth.cnrs.fr, boudjema@lapth.cnrs.fr\\}

\author{A. Pukhov $^*$}

\address{Skobeltsyn Institute of Nuclear Physics, Lomonosov Moscow State University,\\
Leninskie gory, GSP-1, Moscow, 119991, Russia\\
$^*$E-mail: pukhov@lapth.cnrs.fr\\
www.sinp.msu.ru}

\begin{abstract}
These  lecture notes  describe the micrOMEGAs code for the
calculation of Dark Matter observables in extensions of the standard model.
\end{abstract}


\bodymatter

\section{Introduction}
Since the 1930's, with the pioneering work of Zwicky~\cite{Zwicky:1937zza} on the dynamics of the Coma galaxy and the observation~\cite{Rubin:1970zza}
thirty years later of the flatness  of the rotation curves of spiral galaxies, 
evidence for the existence of some missing non luminous matter 
has been steadily gathering. The last two decades or so have witnessed spectacular advances in cosmology and astrophysics confirming  that ordinary matter is a minute part of what constitutes the Universe at large. Most spectacular has been the study of the Cosmic Microwave Background (CMB), in particular combining the results of the 7-year WMAP data~\cite{Jarosik:2010iu}
on the 6-parameter $\Lambda$CDM model, the baryon acoustic oscillations from SDSS\cite{Percival:2009xn} and the most recent determination of the Hubble constant~\cite{Riess:2009pu} one~\cite{Komatsu:2010fb} arrives at a measurement of the relic density to better than $3\%$. 
Yet the exact nature of this dark matter and its microscopic 
properties remain mysterious. At the same time the field of high energy physics has been rich in discoveries of a large number of particles.  
All these  particles can in fact fit very neatly with a modicum of elementary building blocks within the much successful standard 
model ({\bf SM}). Were it not for the masses 
of these particles the dynamics of the standard model would  require less than a handful of parameters which makes the theory very predictive. Yet, none of the particles of the SM contributes much to the weight of the Universe, therefore 
Dark Matter ({\bf DM}) is certainly New Physics. 
Moreover, the problem of mass in the SM is also still mysterious. Electroweak symmetry breaking and  the mechanism behind the generation of mass need elucidation. The SM description poses serious conceptual problems having to do with the missing scalar particle of the SM: the Higgs particle. At the heart of the problem, the naturalness problem, is the observation that there is no symmetry to protect  the mass of a lone elementary scalar like the SM Higgs. This fact has been behind the intense activity in the construction of New Physics models.  Until a few years ago, the epitome of this New Physics has been supersymmetry which when endowed with a discrete symmetry, called R-parity, furnishes a good dark matter candidate. Recently, a few alternatives for the New Physics  have been put forward. Originally, they were confined to solving the Higgs problem, but it has been discovered that, generically, their most viable implementation (in
accord with electroweak precision data, proton decay, {\it etc.}) fares far better if a discrete symmetry is embedded in the model. 
The discrete symmetry is behind the existence of a possible dark matter candidate.
We will call  {\it even} the particles which are  neutral  with respect to the
symmetry and {\it odd} the ones which  get non-trivial factors\footnote{
This terminology comes from the widely used $Z_2$ case}. 
All the SM particles share the same quantum number (even) which sets them apart from   
most of the New Physics particles which have a non even quantum number. 
This makes the lightest New Physics particle with this non even  quantum number 
a stable particle
 which is, beside its electrically neutral character, a potentially good dark matter
candidate.
 Among the most popular possibilities, let us mention some of the candidates and the discrete  symmetry behind each of these candidates. 
\begin{itemlist}
\item R-parity (a $Z_2$ symmetry)\cite{Martin:1997ns} with the  DM in Supersymmetry which is a Majorana fermion
\item KK parity (a $Z_2$ symmetry with $x_5 \to -x_5$)\cite{Cheng:2002ab} and the DM in Universal Extra Dimension which is a  gauge boson
\item T-parity (a $Z_2$ symmetry)\cite{Cheng:2004yc} in Little Higgs model with the DM which is a gauge boson
\item  $Z_3$ symmetry in warped GUTs with the DM\cite{Agashe:2004ci}   which is  a fermion
\end{itemlist}
 Models with $Z_n$  symmetry and $n>3$ have  interesting
phenomenology\cite{Belanger:2012vp} since there can be more than one stable Dark Matter particle.

Therefore,  with the fact that a very large class of models for the New Physics whose primary aim is a better description of the Higgs sector of the SM provide, as a bonus, a  candidate for DM, it is fair to say that we are witnessing the emergence of a strong cross breeding between high energy collider physics on the one hand and cosmology and astrophysics on the other to unravel the mystery of DM. This should be set in the new landscape where a wealth of data and analyses  are being conducted at the colliders, in particular the LHC, as well as important non collider experiments in astrophysics and cosmology. At the {\em cosmological level}, weighing the Universe will be achieved with even higher precision with Planck~\cite{Ade:2011ah}. {\em Direct detection} of Dark Matter in underground experiments, where one measures the recoil of a nucleus due to the Dark Matter particle impinging on it  is being carried out by many collaboration using complementary techniques and nuclear material
DAMA\cite{Bernabei:2010mq},  CDMS\cite{Ahmed:2010wy},
XENON\cite{Aprile:2011hi}. Many {\em indirect detection} experiments are also at work gathering signals from the annihilation of Dark Matter that takes place for example in the galactic halo. These result in  fluxes of  $\gamma$,
$e^{\pm}$, $p^{\pm}$ and neutrinos which can reach the Earth.  Photons and
neutrinos propagate  directly, but the charged particles path and their energy spectra get distorted  by the magnetic fields.  These signals would be detected by satellites
and ground experiments such as  Pamela\cite{Adriani:2008zr,Adriani:2008zq}, HEAT\cite{Beatty:2004cy},
Fermi\cite{Abdo:2009zk,Meurer:2009ir}, ATIC\cite{Chang:2008aa},
HESS\cite{Aharonian:2009ah,Strong:2005zx,Aharonian:2006au},
EGRET\cite{Thompson:2008rw}, INTEGRAL\cite{Strong:2005zx}. New data  on $\gamma,e^\pm, p^\pm$ from
AMS02\cite{Aguilar:2002ad} experiment are expected soon. Other types of experiments are dedicated to analysing the Dark Matter neutrinos  as they get accumulated in the core of the Sun or the Earth.
Super-Kamiokande\cite{Feng:2008qn,Desai:2004pq} and IceCube~\cite{Abbasi:2011eq} are two such neutrino observatories. Reconstructing the microscopic properties of Dark Matter at the LHC and future linear colliders could provide invaluable input for direct and indirect experiments as well as cosmology since this will allow access to a better understanding of the density distribution of dark matter as well as their velocity distribution.

Simulation and Monte Carlo codes for BSM physics at colliders have been around for quite some time. Automatic codes for the generation of matrix elements and cross sections for the colliders are now also numerous. At the colliders the initial state consisting of SM particles is well defined even if in the case of hadronic machines one needs a convolution over structure functions.   Predictions are then made for a specific channel or a BSM particle in the final state. The task 
of a DM code that returns the value of the relic density requires the calculation of a very large number of channels and processes. First of all one needs to identify 
what could be a potentially valid DM candidate (neutrality and stability are a first requirement). Once this is set one generally needs to calculate a large number of processes corresponding to the annihilation of this candidate to all possible SM final states. In general,  since the BSM model has not been constrained and its parameters not measured one has to allow for the calculation of a very large number of processes depending on the properties of the DM particle. In the MSSM for example, one has to be ready to calculate the rates for some 3000 processes. Early codes for the calculations of the relic density listed only a few processes 
that were, at some stage, of a particular interest. Whenever a new mechanism was deemed interesting new calculations were added. 
Indirect detection codes require the decay and fragmentation products of the SM produced in annihilation of DM particles. Furthermore sophisticated modeling of the propagation of  charged particles is needed. In direct detection the rates have to be parametrized and evaluated at very small recoil energy, interaction with nuclei require elements from nuclear physics for instance (form factors,..). All these different observables need to be ``convoluted" with different  DM density distributions and call for a knowledge of the velocity distribution. For the relic density a model of cosmology has to be invoked to take into account the evolution of the Universe, the dilution of the DM and its decoupling. \\

{\tt micrOMEGAs} has been developed with the aim of providing  the value the relic density, the 
fluxes of photons, antiprotons, and positrons for indirect
DM searches; cross sections of DM interactions with
nuclei and energy distribution of recoil nuclei; neutrino and the corresponding muon flux
from DM particles captured by the Sun; collider cross
sections and partial decay widths of particles within a BSM that provides a possible WIMP (weakly Interacting Massive Particle) DM candidate.  What sets {\tt micrOMEGAs} apart from other codes is its ability, once given a {\em Model File} that encodes a BSM model, to output, for any set of parameters of the model, all the observables we have just listed.  All the needed cross sections are built up on the fly. There are several packages which
 calculate different properties of DM  within the  very popular minimal supersymmetric standard  model (MSSM): DarkSUSY\cite{Gondolo:2004sc}
SuperIso\cite{Arbey:2011zz} and Isared\cite{Baer:2004qq}.  The modular structure of {\tt micrOMEGAs} with the (self) automatic generation of all the needed matrix elements and cross sections allows {\tt micrOMEGAs} to tackle practically any model.  The package has been developed  within a French-Russian
 collaboration by G.B\'elanger,  F.Boudjema (LAPTh),  A. Pukhov (SINP), and  A.Semenov (JINR).
 The various features of the {\tt micrOMEGAs} code are described in a series of  papers \cite{Belanger:2001fz,
 Belanger:2004yn,Belanger:2006is,Belanger:2008sj,Belanger:2010gh}. In these
 lectures we describe the micrOMEGAs package, give the main formulas related to the
 calculation of DM relic density and DM signals and present some examples of
{\tt micrOMEGAs} output.  The {\tt micrOMEGAs} package is
 accompanied with an on-line manual which provides a detailed description of  all functions included in the
 package. The user can refer to this manual for a complete specification of the functions and more detailed  information on the program.

For a review on Dark Matter
see\cite{Bertone:2010zz,Bertone:2004pz,Jungman:1995df}.
 
\section{An overview of the modules of {\tt micrOMEGAs}}
\begin{figure}
\begin{center}
 \includegraphics[height=\textwidth, width=0.8\textwidth,angle=-90]{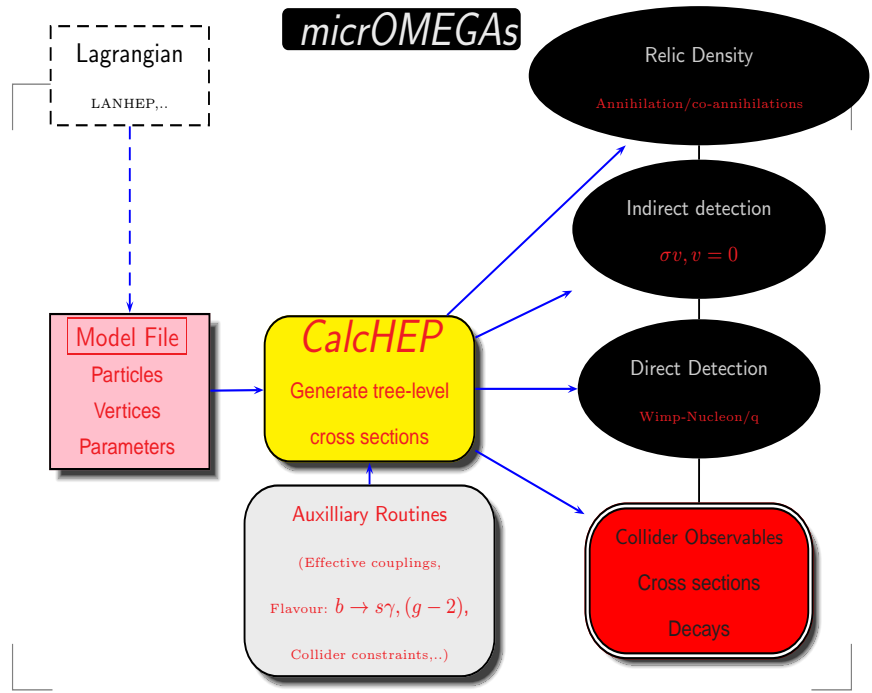}
 \caption{ The micrOMEGAs flow chart.}
\label{fig:micromegas_chart}
\end{center}
\end{figure}

The chart flow of {\tt micrOMEGAs}~\cite{Belanger:2001fz,
 Belanger:2004yn,Belanger:2006is,Belanger:2008sj,Belanger:2010gh} is displayed in Fig.~\ref{fig:micromegas_chart}

All the observables we have pointed at require the computation of interaction rates. The calculation of the cross sections in different kinematical regimes is at the heart of the system. We rely on {\tt CalcHEP}\cite{Pukhov:2004ca,Pukhov:2012} for the generation of all tree-level cross sections. Naturally a model must be defined.  {\tt CalcHEP} requires therefore a model file which defines the nature of all particles in the model (spin, charges,..), parameters (masses, couplings) as well as the interaction vertices or in other words the Feynman rules. Once this is specified in the proper format,  {\tt CalcHEP}  proceeds to the identification of the DM particle. The $Z_n$ assignment is therefore crucial as is of course the mass ordering and the electric neutral character of the WIMP candidate. The current public  version of  micrOMEGAs  can only treat models   with $Z_2$ and $Z_3$  discrete symmetries.
The code is then ready to process any needed cross section. For some models, for example the MSSM, deriving the Feynman rules is a horrendous task. Traditionally 
{\tt micrOMEGAs} has relied on {\tt LanHEP}~\cite{Semenov:2008jy} which is  a code that outputs the model file once given 
the Lagrangian. {\tt FEYNRULES}~\cite{Christensen:2008py} is another recent code that can achieve the same effect. From the cross sections which are model specific, the code calls different shared libraries (common to all models) to output the value of 
\begin{itemlist}
  \item the relic density  
  \item  the rates for indirect detection of $e^+, \bar{p}, \gamma, \nu$. For the case of $\nu$ this includes capture by the Sun and the Earth.  
  \item direct detection for specific targets in large-scale underground experiments.
  \item cross sections at colliders and branching ratios for various particles of the model.  
 \end{itemlist}
This sequence of calls and computations is automated. The code includes also many auxiliary routines which are model specific. Since {\tt CalcHEP} is a tree level cross section generator some important radiative corrections must be introduced through the auxiliary routines. This is the case of the MSSM where the mass of the lightest Higgs must be corrected in a coherent way through an effective Lagrangian that can be interfaced to some spectrum calculator, for example {\tt FeynHiggs}~\cite{Heinemeyer:1998yj} or corrections to Higgs couplings ({\tt HDECAY}~\cite{Djouadi:1997yw}). Input of the MSSM mass spectrum is also implemented through SLHA\cite{Skands:2003cj,Allanach:2008qq}. Other routines include computations such as $(g-2)$, $b \to s \gamma$, $B_s \to \muˆ+ \muˆ-$,..Bounds on some masses and parameters can be easily 
input by the user with sometimes the help of external codes such as {\tt HiggsBounds}\cite{Bechtle:2008jh,Bechtle:2011sb}. 
The code is ``open source" and allows to add a large number of models and also different external codes for some of the auxiliary routines.

%

\section {Downloading and compilation of micrOMEGAs.}
To   download  micrOMEGAs, go to    \\  
\verb|     http://lapth.in2p3.fr/micromegas|\\
and unpack the file received, \verb|micromegas_|\VERSION\verb|.tgz|, with the command\\
\verb|     tar -xvzf micromegas_|\VERSION\verb|.tgz|\\
This should create the directory \verb|micromegas_|\VERSION\verb|/| which occupies about 40
Mb of disk space. You will need more disk space after compilation of
specific models and generation of matrix elements.
In case of problems and questions\\
\verb|     email: micro.omegas@lapp.in2p3.fr|\\

\subsection{File structure of micrOMEGAs.}
\label{file_structure}
\verb|Makefile            |  to compile the kernel of the package               \\
\verb|manual|\VERSION\verb|.tex      |   the manual: description of micrOMEGAs routines \\
\verb|CalcHEP_src/        |        generator of matrix elements for micrOMEGAs    \\
\verb|sources/            |        micrOMEGAs code                               \\
\verb|newProject          |     to create a new model directory   structure                           \\
{\it MSSM model directory}                                                          \\
\verb|MSSM/               |                                                      \\
\verb|   Makefile         |  to compile the code and executable for  this model \\
\verb|   main.c[pp] main.F|       files with sample {\it main} programs      \\
\verb|   lib/             |      directory for routines specific to this model   \\
\verb|       Makefile     |   to compile auxiliary code library {\it lib/aLib.a}    \\
\verb|       *.c *.f      |      source codes of auxiliary functions             \\
\verb|   work/            |              CalcHEP working directory for thegeneration of   \\
\verb|                    |             matrix elements                                    \\
\verb|       Makefile     |  to compile library {\it work/work\_aux.a}           \\ 
\verb|       models/      | directory for files  which specifies the model\\ 
\verb|         vars1.mdl  |  free  variables   \\
\verb|         func1.mdl  |  constrained variables   \\
\verb|         prtcls1.mdl|  particles  \\
\verb|         lgrng1.mdl |   Feynman rules\\
\verb|       tmp/         | auxiliary directories for CalcHEP sessions    \\
\verb|       results/  |                                                  \\
\verb|       so_generated/|   storage  of  matrix elements generated by CalcHEP \\
\verb|    calchep/        |   directory for interactive CalcHEP sessions    \\
{\it Directories of other models which have the same structure as} {\tt  MSSM/ }\\
\verb| NMSSM/             |         Next-to-Minimal SUSY Model\cite{Ellwanger:2006rn,Belanger:2005kh} \\
\verb| CPVMSSM/           |         MSSM with complex parameters\cite{Lee:2003nta,  Belanger:2006qa} \\
\verb| IDM/               |         Inert Doublet Model\cite{Honorez:2010re}  \\
\verb| LHM/               |         Little Higgs Model\cite{Belyaev:2006jh} \\
\verb| RHNM/              |         Right-handed Neutrino Model\cite{Belanger:2007dx}                  \\
\verb| etc/               |          for testing                                 \\

\subsection{ Compilation of CalcHEP and micrOMEGAs routines.}
   CalcHEP and micrOMEGAs are compiled by {\it gmake}. Go to the micrOMEGAs directory
and launch\\
\verb|     gmake|\\
If {\tt gmake} is absent, then {\tt make} should work like {\tt gmake}.
In principle  micrOMEGAs  defines automatically the names of {\it C} and {\it
Fortran} compilers and the flags for
compilation. If you meet a  problem, open the file which contain the compiler specifications, 
\verb|CalcHEP_src/FlagsForSh|,
 improve it, and launch {\tt [g]make} 
again. The file  is written is {\bf sh} script format and looks like
\begin{verbatim}
         # C compiler
         CC="gcc"
         # Flags for C compiler
         CFLAGS="-g -fsigned-char"
         # Disposition of header files for X11
         HX11=
         # Disposition of lX11
         LX11="-lX11"
         # Fortran compiler
         FC="gfortran"
         FFLAGS="-fno-automatic"
         ........
\end{verbatim}
After a successful definition of compilers and their flags,   micrOMEGAs rewrites the file 
 {\it FlagsForSh} into {\it FlagsForMake} and substitutes its contents in all {\it
Makefile}s of the package.\\
\verb|     [g]make clean|    deletes all generated files, but asks permission to
delete {\it FlagsForSh}.\\
\verb|     [g]make flags|       only generates FlagsForSh. It allows to check and
change  flags before compilation of codes.


\subsection{Module structure of main programs.}
Each model included in micrOMEGAs  is accompanied with sample files for
C and Fortran programs which call micrOMEGAs routines, the {\it main.c}, {\it main.F} files.  
These files   consist of
several modules enclosed between the instructions
\begin{verbatim}
#ifdef XXXXX
  ....................
#endif
\end{verbatim}
Each of these blocks  contains some code for a specific problem
{\small
\begin{verbatim}
#define MASSES_INFO        //Displays information about mass spectrum 
#define CONSTRAINTS        //Displays B_>sgamma, Bs->mumu, etc
#define OMEGA              //Calculates the relic density 
#define INDIRECT_DETECTION //Signals of DM annihilation in galactic halo
#define RESET_FORMFACTORS //Redefinition of Form Factors and other
                           //parameters 
#define CDM_NUCLEON        //Calculates amplitudes and cross-sections
                           //for DM-nucleon collisions 
#define CDM_NUCLEUS       //Calculates number of events for 1kg*day
                         //and recoil energy distribution for various nuclei
#define NEUTRINO         //Calculates flux of solar neutrinos and
                         //the corresponding muon flux 
#define DECAYS           //Calculate decay widths and branching ratios  
#define CROSS_SECTIONS   //Calculate cross sections 
\end{verbatim}
}
All these modules are completely independent. The user can comment or
uncomment any set of {\it define} instructions to suit his/her need. 

\subsection{Compilation of codes for specific models.}
 After compilation of micrOMEGAs one has to compile
the executable to compute DM related observables in a specific model. To
do this, go to the model directory, say MSSM,  and launch\\
\verb|     [g]make main=main.c|\\
It should generate the executable {\tt main}. In the same manner\\
\verb|     gmake main=|{\it filename}.{\it ext}\\ 
generates the executable {\tt filename}  based on the source file {\it
filename.ext}.
For {\it ext}  we support 3 options: {\it 'c'} , {\it 'F'}, {\it 'cpp'} which correspond to
{\tt C}, {\tt FORTRAN} and {\tt C++} sources.
{\tt [g]make} called  in the model directory automatically  launches {\tt [g]make}
in subdirectories {\it lib} and {\it work} to compile \\
 \verb|     lib/aLib.a|   - library of auxiliary model functions, e.g. constraints,\\
 \verb|     work/work_aux.a| - library of model particles, free and dependent parameters.\\

\subsection{ Command line parameters of main programs.}
\label{sec:command}
Default versions of {\it main.c/F}  programs need some arguments
which have to be specified in command lines. If launched without
arguments {\it main} explains which parameter are needed. 
As a rule  {\it main}  needs  the name of a file containing the
numerical values of the free parameters of the file. The structure of a file
record should be\\
\verb|Name       Value # comment ( optional)|\\
For instance, an Inert Doublet model (IDM) input file, see  section\ref{sec:IDM}, contains
\begin{verbatim}
 Mh    125   # mass of SM Higgs 
 MHC   200   # mass of charged Higgs ~H+
 MH3   200   # mass of odd Higgs ~H3
 MHX    63.2 # mass of ~X particle
 la2  0.01   # \lambda_2  coupling
 laL  0.01   # 0.5*(\lambda_3+\lambda_4+\lambda_5)
\end{verbatim}

In other cases, different inputs can be required. For example, in the MSSM with input parameters defined at the GUT scale,
the parameters have to be provide in a command line. Launching \verb|./main| will return 
\begin{verbatim}
      This program needs 4 parameters:
     m0      common scalar mass at GUT scale
     mhf common gaugino mass at GUT scale
     a0     trilinear soft breaking parameter at GUT scale
     tb    tan(beta)
   Auxiliary parameters are:
     sgn +/-1, sign of Higgsino mass term (default 1)
     Mtp top quark pole mass
     MbMb Mb(Mb) scale independent b-quark mass
     alfSMZ strong coupling at MZ
   Example: ./main 120 500 -350 10 1 173.1
\end{verbatim}

\section{Structure of model files in CalcHEP}
\label{sec:structure}

micrOMEGAs uses CalcHEP\cite{Pukhov:2004ca,Pukhov:2012} to generate automatically the code for the computation of matrix
elements. It makes micrOMEGAs flexible for new model implementation.
The complete manual for  CalcHEP  and in particular
a description of the structure of CalcHEP model files can be found at\\
\verb|     http://theory.sinp.msu.ru/~/pukhov/calchep.html|\\
Here we present only the information about model files  needed for understanding  and using
 micrOMEGAs. As an example, we will use here the Inert Doublet model.
 
 \subsection{A simple example, the Inert Doublet Model.}
\label{sec:IDM}

The IDM\cite{Barbieri:2006dq} contains two $SU(2) \times U(1)$ scalar doublets. In the unitary
gauge 
$$ H_1=\left( \begin{array}{c}
  0\\
  \langle v \rangle + h/\sqrt{2}
\end{array}\right)\;\;,\;\;
 H_2=\left( \begin{array}{c}
  \widetilde{H}^+\\
  (\widetilde{X}+i\widetilde{H}_3)/\sqrt{2}
\end{array}
\right)
$$
where $H_1$ is the SM Higgs doublet and $H_2$ is a new {\it inert} doublet 
which does not couple to quarks and leptons and is odd under a new $Z_2$ symmetry. Unlike the SM scalar doublet it does 
not develop a vacuum expectation value.
$\widetilde{H}^+$, $\widetilde{X}$, and $\widetilde{H}_3$ are the new fields
of the model. The IDM Lagrangian contains only even powers of the doublet $H_2$
\begin{eqnarray}
   {\cal L}&=&(SM\;terms)+ D^{\mu}H_2^*D_{\mu}H_2 -\mu^2 |H_2^2|^2  \\
   \nonumber
   && -\lambda_2 H_2^2 -\lambda_3 H_1^2H_2^2
 -\lambda_4 |H_1^*H_2|^2  - \lambda_5 Re[(H_1^*H_2)^2]
\end{eqnarray}
 Because of the   $H_2 \to -H_2$ symmetry,   the lightest   new particle  is stable. 
In the following example we will use the masses of new particles as well as 
$ \lambda_2$ and $\lambda_L=(\lambda_3+\lambda_4+\lambda_5)/2$ as independent
parameters,
 The couplings  $\mu$, $\lambda_3$, $\lambda_4$, $\lambda_5$  
can be expressed in terms of the independent parameters. 
For  details  of IDM and further references see  Ref.\cite{Honorez:2010re}. 

\subsection{ Particles of the model.}
The list of  particles is given  in the file {\it models/prtcls1.mdl}, for example \
\begin{verbatim}
Inert Doublet
Full    Name | P | aP|number|spin2|mass|width|color|aux|> LaTeX(A)
photon       |A  |A  |22    |2    |0   |0    |1    |G  |A
Z boson      |Z  |Z  |23    |2    |MZ  |wZ   |1    |G  |Z
gluon        |G  |G  |21    |2    |0   |0    |8    |G  |G
W boson      |W+ |W- |24    |2    |MW  |wW   |1    |G  |W^+
neutrino     |n1 |N1 |12    |1    |0   |0    |1    |L  |\nu^e
electron     |e1 |E1 |11    |1    |0   |0    |1    |   |e
mu-neutrino  |n2 |N2 |14    |1    |0   |0    |1    |L  |\nu^\mu
muon         |e2 |E2 |13    |1    |Mm  |0    |1    |   |\mu
tau-neutrino |n3 |N3 |16    |1    |0   |0    |1    |L  |\nu^\tau
tau-lepton   |e3 |E3 |15    |1    |Mt  |0    |1    |   |\tau
u-quark      |u  |U  |2     |1    |0   |0    |3    |   |u
d-quark      |d  |D  |1     |1    |0   |0    |3    |   |d
c-quark      |c  |C  |4     |1    |Mc  |0    |3    |   |c
s-quark      |s  |S  |3     |1    |Ms  |0    |3    |   |s
t-quark      |t  |T  |6     |1    |Mtop|wtop |3    |   |t
b-quark      |b  |B  |5     |1    |Mb  |0    |3    |   |b
Higgs        |h  |h  |25    |0    |Mh  |!wh  |1    |   |h
odd Higgs    |~H3|~H3|36    |0    |MH3 |!wH3 |1    |   |H_3
Charged Higgs|~H+|~H-|37    |0    |MHC |!wHC |1    |   |H^+
second Higgs |~X |~X |35    |0    |MHX |!wHX |1    |   |X
\end{verbatim}
The columns of this table have the following meaning:
\begin{itemlist}
\item{\bf P, aP -} names of particle and antiparticle used in  interaction vertices.
\item{\bf number -}  Particle number according to the Monte Carlo numbering
scheme\cite{Nakamura:2010zzi} (PDG code).
For new particles not listed in PDG, one can use any number as long as it is not already used for some 
particle. In order to recognize which name corresponds to a given particle, micrOMEGAs uses PDG numbers. 
Interface with other packages is also
based on these codes.
\item{spin2 -} twice the  particle spin
\item{mass -}  mass identifier
\item{width -} width identifier, when an exclamation mark precedes  the width
symbol, this width will be calculated automatically. 
\item{color -} dimension of representation of color group.
\item{aux -} 'G' means that this particle is a gauge boson which is treated in the t'Hooft-Feynman
gauge. Interactions of ghost fields have to be specified in the file for vertices and the sum over ghost fields
will be taken into account automatically. 'L/R'  denotes  massless fermions which exist only
in left/right states. A number written in this column means $3Q$ where $Q$ is the electric charge. 
This charge is usually detected by CalcHEP automatically from the interaction vertices, in  cases where it is not possible, the number
should be specified in this column. 
\item{Latex(A)} Particle names in LaTex format (optional)
\end{itemlist}

Note that the names of particles in the odd sector should started with tilde \verb|~|.

\subsection{Free parameters of the model.} 
\label{sec:free_parameters}
The independent (free) parameters of the model are listed  in the file {\it
models/vars1.mdl}~\footnote{See micrOMEGAs file structures in section
\ref{file_structure}}. For example, 
\begin{verbatim}
Inert Doublet Model
 Variables
  Name | Value     | Comment                          |
EE     |0.31333    |Electromagnetic coupling constant
SW     |0.474      |sin of the Weinberg angle
MZ     |91.187     |Mass of Z
MHX    |63.2       |Mass of Inert Doublet Higgs
MH3    |200        |Mass of CP-odd Higgs
MHC    |200        |Mass of charged Higgs
Mh     |125        |Mass of standard Higgs
LaL    |0.01       |Coupling in Inert Sector
La2    |0.01       |Coupling in Inert Sector
..................................
\end{verbatim}
Three functions can be used to extract or set the values of the free parameters:\\
\verb|     findValW(char*name)| - returns the numerical value of parameter {\it name}.\\
\verb|     assignValW(char*name,double value)| - assigns new value to parameter {\it name}.
It gives a warning if {\it name} does not correspond to any parameter.\\
\verb|     readVar(fileName)| - 
allows to  download a set of parameters, e.g. from a file. The structure of the file
is explained in section \ref{sec:command}. \verb|readVar| returns zero when the file has been read successfully
or a positive value  corresponding to the line in the file which contained an error. 

\subsection {Constrained parameter of the model.}

Constrained parameters are stored in the file {\it models/func1.mdl}. For example
\begin{verbatim}
  Inert Doublet Model
   Constraints
   Name |> Expression
  CW    |sqrt(1-SW^2)             % cosine of Weinberg angle
  MW    |MZ*CW                    % W+ mass   
  Mb    |MbEff(Q)                 % b-quark effective mass
  Mc    |McEff(Q)
  mu2   |MHX^2-laL*(2*MW/EE*SW)^2        % mu^2
  la3   |2*(MHC^2-mu2)/(2*MW/EE*SW)^2    % \lambda_3 
  la5   |(MHX^2-MH3^2)/(2*MW/EE*SW)^2    % \lambda_5
  la4   |2*laL-la3-la5                   % \lambda_4   
 %Local!|
  ..................
\end{verbatim}
The constrained parameters are divided into two parts. The {\it public} parameters
listed at the top of the file   and the {\it local} parameters at the bottom of the 
file. The splitting of parameters into two sets is constructed using the following rules: {\it public} parameters include {\bf i)} all parameters involved in the
calculation of particle masses and  {\bf ii)} all parameters which call external functions for calculations.  
The value of {\it public} parameters  can be obtained  via  the function
\verb|findValW(name)|.

 Local parameters appear only
inside the  matrix elements codes and can not be accessed with  micrOMEGAs. 
The user can enlarge the list of {\it public}
parameters by writing a record in func1.mdl  which starts with 
\verb|%Local!|. 
Then all parameters disposed above this record become {\it
public}.
This separation of parameters into two classes is needed for optimization of
calculations in models where there is a large  number of constrained parameters.

The calculation of all {\it public} model constraints and of the particle spectrum is done with\\
\verb|     sortOddParticles(text)| ,\\
which computes all constraints, sorts the odd particles with increasing masses, writes the name of the 
lightest odd particle in {\it text} and assigns its mass to the global parameter {\tt Mcdm}. This routine returns a non zero error code if some constraint cannot be calculated. The name of the corresponding constraint is written in {\it text}. This routine has to be called before any other functions when starting to work with a model, furthermore it has to be called after a reassignment of any input parameter.

The properties of particles can be tested with the two functions\\
\verb|     pMass(pName)|, which gives the mass of particle {\it pName}.\\
\verb|     qNumbers(pName, &spin2,&charge3,&cdim)|,
which calculates the quantum numbers of a particle and returns directly its
{\it PDG} code. This function  allows  to  check that the Dark Matter candidate has no electric or color charges.

The  table of Feynman rules for each model is rather long. For example, one can check 
the {\it IDM/work/models/lgrng1.mdl} file to see how the interactions of SM and Inert doublet 
particles are  presented in CalcHEP/micrOMEGAs.

\subsection{ Example: MSSM mass spectrum.}
\label{MSSM_example}
As an example we present here the code used for finding the properties of the DM candidate 
in the  MSSM model with parameters defined at GUT scale. This option is specified by\\ 
\verb|     #define SUGRA|\\ at the top of  the main file in the MSSM directory.
{\small
\begin{verbatim}
// CODE 
err=sortOddParticles(cdmName); 
if(err){ printf("Can't calculate %s\n",cdmName); return 1;}
qNumbers(cdmName,&spin2, &charge3, &cdim);
printf("\nDark matter candidate is '%s' with spin=%d/2  mass=%.2E\n",
   cdmName,spin2,Mcdm); 
\end{verbatim}
} 
\noindent
With the following input parameters (in GeV for dimensionful variables)
$$   M_0=120,\; M_{1/2}=500,\;   A_0=-350,\; \tan\beta=10,\;
sign(\mu)=1,\;   M_{top}=173.1;$$
the micrOMEGAs output is 
\begin{verbatim}
Dark matter candidate is '~o1' with spin=1/2  mass=2.06E+02
\end{verbatim}

One can print  the masses of particles included in the model with the commands\\
\verb|     printHiggs(FD)|-   prints masses and widths of scalar
(Higgs) particles in open file {\it FD}.\\  
\verb|     printMasses(FD,sort)|-  prints masses of {\it odd} sector
particles. If $sort \ne 0$ particles are sorted with respect to their masses. 
The corresponding  micrOMEGAs output for these two functions is 
{\small
\begin{verbatim}
Higgs masses and widths
    h   115.99 3.07E­03
    H   766.59 1.65E+00
   H3   766.34 1.68E+00
   H+   770.85 1.61E+00
Masses of odd sector Particles:
~o1  : MNE1  =   205.9  || ~l1  : MSl1  =   212.3   || ~eR  : MSeR  =223.7
~mR  : MSmR  =   223.7  || ~nl  : MSnl  =   344.8   || ~ne  : MSne  =346.8
~nm  : MSnm  =   346.8  || ~eL  : MSeL  =   355.4   || ~mL  : MSmL  =355.4
~l2  : MSl2  =   356.5  || ~1+  : MC1   =   389.6   || ~o2  : MNE2  =389.7
~o3  : MNE3  =   692.7  || ~2+  : MC2   =   704.2   || ~o4  : MNE4  =704.3
~t1  : MSt1  =   767.9  || ~b1  : MSb1  =   959.3   || ~b2  : MSb2  =1005.7
~t2  : MSt2  =  1006.1  || ~dR  : MSdR  =  1009.9   || ~sR  : MSsR  =1009.9
~uR  : MSuR  =  1013.4  || ~cR  : MScR  =  1013.4   || ~uL  : MSuL  =1050.2
~cL  : MScL  =  1050.2  || ~dL  : MSdL  =  1053.1   || ~sL  : MSsL  =1053.1
~g   : MSG   =  1146.4  ||
\end{verbatim}
}

\section{ Runtime generation of matrix elements by CalcHEP.}
\subsection{ Running CalcHEP in blind mode.}
  In general  the  CalcHEP version  used in micrOMEGAs for generating  matrix elements  
is a menu driven package.  micrOMEGAs users can go to {\it ./calchep}
subdirectory found in   any    {\it MODEL} directory and launch the
{\it ./calchep} command script to see how CalcHEP works and to check
different matrix elements of the DM model used.
However CalcHEP also  works in {\it blind} mode 
 when  special keys passed  in command line are interpreted as  keyboard
hits. For example, the command  \\
\verb|     CalcHEP_src/bin/s_calchep -blind "{{~dm,~dm­>2*x{{{[{[{{0"|\\
generates C-code for annihilation of two DM particles (\verb|~dm|) into 2
arbitrary particles of the model. Here  {\verb|{|} simulates hit of  the {\tt Enter}
key,  {\verb|}| simulates {\tt Esc},  \verb|[| and \verb|]| are used for down
and up arrow keys respectively.

\subsection{ Dynamic linking  and generation of shared libraries of matrix elements.}
  micrOMEGAs transfers the C-code of matrix elements into shared library.
The general form for such command is\\ 
\verb|     gcc -shared -o library_name.so matrix_element_files.c |\\
All shared libraries of matrix elements for a given MODEL are stored in
the directory\\
\verb|     MODEL/work/so_generated/ |\\
They are recompiled automatically if the model is changed.
  Shared library can be linked dynamically in runtime\footnote{See output of {\it
man} unix command  for {\it dlopen}, {\it dlsym}}. A simple example of such code is presented below
\begin{verbatim}
* =========== file dltest.c =========*/
#include <stdio.h>
#include <dlfcn.h>
int main(void)
{ void *handle; double (*arcsin)(double);
  handle = dlopen("libm.so",RTLD_NOW);
  arcsin = dlsym(handle,"asin");
  printf ("%f\n", 2*(*arcsin)(1));
}
\end{verbatim}
Here we open in runtime  the standard library of mathematical functions which was not
linked to the code from the beginning, find an address of the {\it arcsine }
function and calculate $2\arcsin(1) $ which should be equal to $\pi$.
To compile the function presented above use \\
\verb|     gcc dtest.c -ldl|\\
The general scheme for matrix element generation in
micrOMEGAs is the following. The downloaded package  does not contain any code for matrix
elements. When micrOMEGAs needs  some matrix element it first checks
the existence of the corresponding library in the directory {\tt
MODELS/work/so\_generation}. If such library already  exists, it is
linked dynamically. If not, micrOMEGAs calls CalcHEP to generate the C-code for the
requested matrix element, transforms it to a shared library, moves this library
to the \verb|so_generated| directory and loads it dynamically. Because of
this set-up the first call to  micrOMEGAs can take some time. However further calls, even with new parameters,   are executed rather fast.

\section{ Calculation of Relic Density.}
 Density of DM in the Universe  is extracted from measurements of
fluctuations of  CMB temperature by the WMAP\cite{Bennett:2003bz,Spergel:2003cb} experiment. 
$$             \Omega_{dm} h^2 = \rho_{dm}/\rho_{c} = 0.105(8) $$
   where h =0.73(3) defines  the present day Hubble expansion rate
$$           H_0=h\cdot 100km/s/Mpc\;\;\;, $$
$\rho_{c}=10.537h^2 GeV/m^3$  is a critical density corresponding to $H_0$. 
Thus  $\rho_{dm}\approx 1.11 GeV/m^3$. The DM density is usually given  by 
the value $\Omega_{dm} h^2$ to avoid the uncertainty associated with the Hubble rate.

We assume that at large temperatures DM particles are in thermal equilibrium with SM
particles, so that   DM self-annihilation is compensated  by  DM
production in collisions of SM particles. Thus the DM density reaches its equilibrium value, $N_{eq}$, described by the Boltzmann formula,
$$ N \approx N_{eq} \approx e^{-\frac{M_{cdm}}{T}}\;\;.$$
When the DM density becomes small and  self-annihilation stops, the  DM 
density is driven by the expansion of the Universe and decreases slowly while
$N_{eq}$ drops rapidly to zero. So, some {\it relic} DM can be observed. 

If the DM is a WIMP with a mass scale from 1-1000GeV and which interact weakly  with SM particles, then its {\it relic} density is in agreement with the value observed, 
$\Omega h^2\approx 0.1$. It is  usually called the {\it WIMP miracle}.  The temperature where self annihilation stops
 is called the {\it freeze-out} temperature.   

\subsection{ Equations of thermal evolution of DM density}
Here we present the   equations  used for the calculation of $\Omega h^2$ using the   {\tt darkOmega} function in micrOMEGAs, see section~\ref{sec:darkomega}.
Let $N$   designate the total  number density of {\it odd} particles. Then the time
evolution of the DM density reads
$$\frac{dN}{dt} = -3H N -\langle\sigma v\rangle_T (N^2 -N_{eq}^2) $$ 
The first term is caused by Hubble expansion which is a function of the current
matter density $\rho$
$$ H=\frac{1}{M_{Planck}}  \sqrt{\frac{8\pi}{3}\rho}  $$ 
 The second term describes two
processes, DM annihilation into SM particles and its reverse.
If $N=N_{eq} $, these two
processes are in thermal equilibrium.  In general, the rate of a reaction is
defined by the product of the particle velocity with the cross section, $\sigma \cdot v $. 
Because the DM velocity follows a Boltzmann distribution, one has to compute the thermally average cross section, 
$\langle \sigma v\rangle_T$.
The equilibrium density $N_{eq}$ equals the sum of contributions of different {\it
odd} particles
$$ N_{eq} = \sum_{i \in oddParticles} N_{eq}^i\;\;,$$
where
$$N_{eq}^i= n^i\int \frac{dp^3}{(2\pi)^3}\left(
e^{\frac{\sqrt{p^2+M_i^2}}{T}}\pm 1\right)^{-1}
\approx n^i\int \frac{dp^3}{(2\pi)^3}e^{-\frac{\sqrt{p^2+M_i^2}}{T}}\;\;,$$
where $n^i$ denotes the  number of   degrees of freedom of particle $i$.  

The total DM number density is obtained after summing on all different odd particles, $N=\sum N_i$, since all odd particles eventually decay into the DM. One can assume a chemical equilibrium
between different components $N_i$ because of the decays of all odd particles
into the lightest one. Thus, 
$$ N_i(T)/N_j(T)= N_{eq}^i(T)/N_{eq}^j(T) \approx e^{-\frac{M_i-M_j}{T}}$$
Consequently the formula for    $\langle\sigma v\rangle_T$ should contain averaging over 
different  particle of the {\it odd} sector.  
One can find explicit formulas for $\langle\sigma v\rangle_T $  in the
papers\cite{Edsjo:1997bg, Belanger:2006is}.

Let us define  $s$ as the entropy density. Because of entropy conservation
$$ \frac{ds}{dt} = -3H\cdot s$$ 
and we can  replace the equation for the number density by a simpler equation for the abundance  $Y=N/s$
\begin{equation}
\label{abundEQ}
 \frac{dY}{ds}= \frac{1}{3H}\langle\sigma v\rangle_T (Y^2-Y_{eq}(T)^2)=
\frac{M_{Planck}}{\sqrt{24\pi\rho}} \langle\sigma v\rangle_T (Y^2-Y_{eq}(T)^2) 
\end{equation}

\subsection{Thermodynamics of Universe}

The formulas that give the contribution  of one relativistic 
degree of freedom to the  entropy density and  energy  density of the Universe are
rather simple. For bosons  
\begin{equation}
\rho_b=\frac{\pi^2}{30}T^4\;,\;\;\;s_b=\frac{2\pi^2}{45}T^3
\end{equation}
The corresponding  functions for fermions have an additional factor $\frac{7}{8}$.
In the generic case one can write
\begin{equation}
\rho(T)=\frac{\pi^2}{30}T^4g_{eff}(T)\;,\;\;\;s(T)=\frac{2\pi^2}{45}T^3
h_{eff}(T)
\end{equation}
where $g_{eff}(T)$/$h_{eff}(T)$  count the effective numbers of degrees of
freedom\footnote{$g_{eff}(T)$ and $h_{eff}(T)$ are related via condition 
$\frac{d\rho(T)}{dT}=T\frac{ds(T)}{dT}$  }. At freeze-out temperatures, the  contribution of Dark Matter  should be strongly suppressed, only
Standard Model particles contribute to $g_{eff}$ and $h_{eff}$.
\footnote{We assume that at low temperatures when DM 
contributes  significantly  to $\rho$ the formation of Dark Matter is finished and $Y(T)=Const$.} 
At low temperature $(T\ll 100 MeV)$, the  effective number of degree of freedom is 
defined by photons, neutrinos, light leptons.  As the temperature increases,  one 
has to add the quarks degrees of freedom. There is a problem of correct matching 
between these two regions.  micrOMEGAs uses the tabulated functions for $g_{eff}(T)/h_{eff}(T)$ 
obtained in \cite{Olive:1980wz,Srednicki:1988ce}. They are displayed in Fig.\ref{fig:heff}.
\begin{figure}
\begin{center}
 \includegraphics[height=6cm, width=10cm]{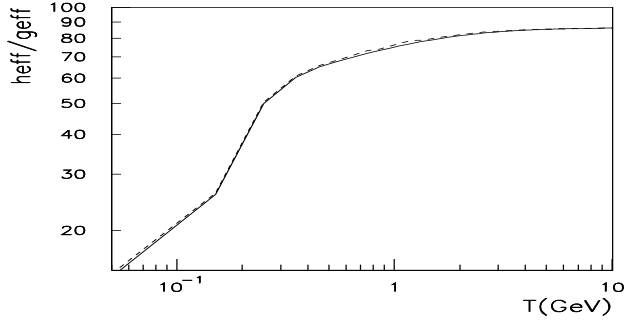}
 \vspace{-.8cm}
 \caption{ $h_{eff}$ (solid)  and $g_{eff}$(dashed)   as  functions of
 temperature.}
\label{fig:heff}
\end{center}
\end{figure}

In order to convert  the abundance $Y_0$ obtained by solving 
Eq.(\ref{abundEQ}) into the DM number density one has to know the value of the entropy today.
The present  day entropy is stored in photons and neutrinos.  Photons have a temperature 
$T=2.725K=2.348\cdot10^{-4}eV=1196 m^{-1}$. When $e^\pm$ pairs
annihilate, their entropy is transformed  into entropy of photons which leads 
to a temperature of photons larger than the  temperature of neutrinos. 
After electrons decouple we have  independent gases of photons and neutrinos.
Counting the degrees of freedom leads to the  entropy density ratio 
$$ \frac{s_{\nu}}{s_{\gamma}} = \frac{6(7/8)}{2+4(7/8)}=\frac{21}{22}$$
which allows to calculate the present day entropy. As a  result   we 
get for the relic density
$$ \Omega h^2 = \frac{M_{cdm}[GeV]}{10.57[GeV/m^3]}   \cdot Y_0 \cdot 2.889\cdot
10^9m^{-3}$$ where $2.889\cdot 10^9m^{-3}$ is the present day
entropy\cite{Nakamura:2010zzi} and  $10.57GeV/m^3$ is the critical density for 
$ H=100\frac{km}{s Mpc}$

\subsection{ Numerical solution of DM thermal evolution equation.}
 \subsubsection{High temperature region}
Equation (\ref{abundEQ}) can be solved numerically by means of the Runge-Kutta
method. The only problem with this method is to find a starting point at
high temperature. Let $\Delta Y = Y-Y_{eq}$. 
Assuming that at high temperature
\begin{equation}
\label{smalldY}
 \Delta Y \ll Y_{eq}\;,\;\;\;  \frac{ d\Delta Y}{ds} \ll \frac{dY_{eq}}{ds}
\end{equation}
we estimate
\begin{equation}
\label{deltaY}
\Delta Y =  \frac{d \log{Y_{eq}(s)}}{ds} \frac{\sqrt{6\pi\rho}}{M_{Planck}\langle\sigma
v\rangle_T}\;\;\;.
\end{equation}
The function $Y_{eq}$ drops rapidly, nevertheless our initial assumption
(\ref{smalldY}) is satisfied because the 
dependence on $Y_{eq}$ is only logarithmic. 
This  gives us a rather simple picture of
freeze-out. We have an almost  constant $\Delta Y$  on the background of a fast
dropping $Y_{eq}$ as the temperature decreases. At some temperature $\Delta Y$ becomes comparable
with $Y_{eq}$ and  we have the freezing-out of $Y\approx \Delta Y$.

micrOMEGAs uses Eq.(\ref{deltaY}) to find the  starting point for the Runge-Kutta
procedure looking for a point where  $\Delta Y \approx 0.1 Y_{eq}$.  

\subsubsection{Freeze-out approximation}
One can use Eq.(\ref{deltaY}) to find $s_f$ ( or the corresponding $T_f$) where
$Y(T_f) \gg Y_{eq}(T)$. micrOMEGAs defines $T_f$ by condition
\begin{equation}
  \label{Tf}
  Y(T_f ) = 2.5 Y_{eq}(T_f)\;\;.
\end{equation}
Below $T_f$  one can neglect the $Y_{eq}^2$ term in 
Eq.(\ref{abundEQ}) and get the explicit solution
\begin{equation}
\label{FO1}
\frac{1}{Y(s_0)} = \frac{1}{2.5Y_{eq}(s_f)} + \int_{s_0}^{s_f} ds
\frac{M_{Planck}\langle\sigma v\rangle_T}{\sqrt{24\pi\rho}} 
\end{equation}  
This solution is called {\it freeze-out approximation} and as a rule works  with
$\approx 2\%$ accuracy.

One can make further approximations and discard the first term in the right side of
(\ref{FO1}). This approximation corresponds to an infinite DM density at $T_f$
and works with $\approx 20\%$ accuracy
\begin{equation}
\label{FO2}
\frac{1}{Y(s_0)} =  \int_{s_0}^{s_f} ds
\frac{M_{Planck}\langle\sigma v\rangle_T}{\sqrt{24\pi\rho}} 
\end{equation}  
micrOMEGAs uses this to  calculate the relative contributions of different channels  to $\Omega^{-1}$. 
It gives an understanding of the physical reactions responsible for relic DM
formation.

\subsection{ The $Z_3$ case.}  
The  $Z_3$ internal charge of particles can be zero, $1$, or $-1$, which
corresponds to phases \{$0,\frac{2\pi}{3},\frac{-2\pi}{3}$\}.   The lightest
particle with non-zero charge has to be stable. Particles with charge $-1$ are
the antiparticles of those with charge $1$. Thus, only one DM candidate is expected.
We have 2 types of reactions which change the DM density
\begin{equation} 
\widetilde{X}, \widetilde{Y} \to SM, SM \;\;\;\;\; \widetilde{X},\widetilde{Y} \to \widetilde{Z}, SM
\end{equation}
The second type of reactions modify the abundance equation,  Eq.(\ref{abundEQ}), 
$$ \frac{dY}{ds}=  \frac{M_{Planck}}{\sqrt{24\pi\rho}}(\langle\sigma v\rangle_T^{11 \to
00}(Y^2-Y_{eq}(T)^2)+0.5\langle\sigma v\rangle_T^{11\to 10}Y\cdot (Y-Y_{eq}(T)) $$
where $\langle\sigma v\rangle_T^{11 \to 00}$, $\langle\sigma v\rangle_T^{11 \to 10}$ define the thermally 
average cross section for 2 particles with charge $\pm 1$ annihilating into  particles with charge 0
and into a charge 1 and a charge 0 respectively.
The modified abundance equation can be solved by the same method as  Eq.(\ref{abundEQ}) and is
included in micrOMEGAs.
The case of  $Z_4$  discrete symmetry  which leads to  
two stable particles with $Z_4$ charges \{1,-1\}  and \{2,-2\}
is not  treated in the  current version of micrOMEGAs, however see Ref.~\cite{Belanger:2012vp} for an application.


\subsection{The micrOMEGAs routines.}
\label{sec:darkomega}
$\Omega h^2$ can be calculated by the function\\
\verb|     omega= darkOmega(&Xf, fast,Beps);| \\
where   {\it  fast} and {\it Beps} are  input parameters.  $fast \ne 0$ leads to an optimized fast
calculation.  The parameter  {\it Besp} allows  to change the number of channels
taken into account  for the evaluation of $\Omega h^2$.
The contribution of channels for which the sum of the mass  of incoming or outgoing
particles, $M_s$, is large is suppressed by the Boltzmann factor 
$$   e^{\frac{2M_{cdm}-M_s}{T_f}} $$ 
where $T_f$ is  the freeze-out temperature (\ref{Tf}). micrOMEGAs discards all channels for
which this factor is smaller then $Beps$. The default value $Beps=10^{-4}$ leads to a robust
evaluation of the relic density. In some special cases, for example, in models with extra dimensions,  there are
too many co-annihilation channels  and it is reasonable to set
$Beps=10^{-2}$ to have a fast calculation. The output parameter 
$$   X_f = M_{cdm}/T_f$$
 characterizes  the freeze-out temperature $T_f$.
To get a reasonable value of the DM density, the   parameter $X_f$ should be about  $25$. 

The micrOMEGAs package contains two routines which are useful for understanding
of relic density formation. \\

\noindent
$\bullet$ \verb|printChannels(Xf,cut,Beps,prcnt,FD)|\\   
writes into opened file \verb|FD| the  contributions  of different channels to $(\Omega
h^2)^{-1}$. Here \verb|Xf| is an input parameter which should
be first evaluated in \verb|darkOmega|. Only  the channels whose
relative contribution is larger than  \verb|cut| will be displayed.
\verb|Beps|
plays the same role as the \verb|darkOmega| routine.
If $prcnt\ne 0$ the contributions are given in percent.
Note that  for this specific purpose  we use the
freeze-out approximation (\ref{FO2}).\\

\noindent
$\bullet$ \verb|vSigma(T,Beps,fast)|\\
calculates the thermally averaged cross section for DM annihilation  times
velocity  at a  temperature T [GeV], see formula (2.6) in Ref.~\cite{Belanger:2001fz}. The
value for $\sigma_v$
is expressed in [pb]. The parameters $Beps$ and  $fast$ work in the same way
as in the \verb|darkOmega| function. In the $Z_3$ case, \verb|vSigma| returns
$$\langle \sigma v \rangle_T^{DM,DM->SM,SM} + 0.5\langle \sigma v \rangle_T^{DM,DM->DM,SM}$$

The micrOMEGAs code contains  the obsolete
\verb|darkOmegaFO| function which calculates $\Omega h^2$ in freeze-out  approximation
(\ref{FO1}).

\subsection{Example of output}
The output  of \verb|darkOmega|, \verb|vSigma| and \verb|printChannels|
commands for the MSSM point of section {\ref{MSSM_example} is presented below,
\begin{verbatim}
==== Calculation of relic density =====
Xf=2.65e+01 Omega=1.10e­01
Channels which contribute to 1/(omega) more than 1%.
Relative contributions in % are displyed
 29% ~o1 ~l1 ­>A l
 21% ~l1 ~l1 ­>l l
  8% ~o1 ~l1 ­>Z l
  6% ~l1 ~L1 ­>A A
  4% ~o1 ~o1 ­>l L
  3% ~o1 ~o1 ­>m M
  3% ~o1 ~o1 ­>e E
  3% ~o1 ~eR ­>A e
  3% ~o1 ~mR ­>A m
  3% ~l1 ~L1 ­>A Z
  3% ~eR ~l1 ­>e l
  3% ~mR ~l1 ­>m l
\end{verbatim}

One can use the \verb|vSigma| function  to study the  dependence of $\langle\sigma v\rangle$ 
on the   temperature, see Fig.\ref{fig:vsigma}.

\begin{figure}
\begin{center}
\vspace{-.8cm}
 \includegraphics[height=7.8cm, width=10cm,angle=0]{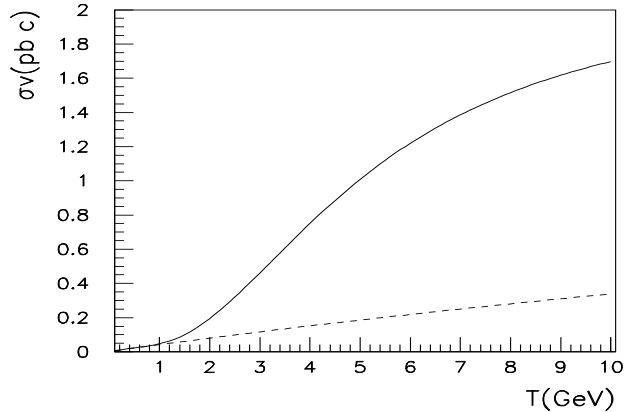}
 \vspace{-1.4cm}
 \caption{$\sigma_v(T)$ for the MSSM point of section \ref{MSSM_example}
 including all coannihilation channes(solid) and the same for only
 neutralino self-annihilation(dashed).}
\label{fig:vsigma}
\end{center}
\end{figure}

For this test point, the contribution of DM (the neutralino - \verb|~o1| ) self- annihilation to $\Omega^{-1}$ is about 10\%. The spectrum also contains 
the {\it stau} particle (\verb|~l1| - superpartner of $\tau$) 
which has a  large self-annihilation cross section as well as a large annihilation cross
section with DM. Furthermore the {\it stau} has
 a small mass difference with the DM
($\approx 6$ GeV).  At the  freeze-out temperature ($T_f\approx 8{\rm GeV}$)   the  {\it stau} density is hardly   
suppressed by the Boltzmann factor, therefore this particle give a large contribution to  $\langle \sigma_v \rangle_T$. Fig.\ref{fig:vsigma}
shows  that at low temperatures where the {\it stau} contribution is
suppressed, the total cross section is small, but at  $T=T_f$ it increases to the
[pb] level, this is sufficient to get a reasonable value for the relic density.

\section{Dark Matter distribution in Milky Way.}
Fluctuations of CMB  temperatures provide crucial  information about DM in the early 
Universe before recombination. Other observables are related to DM in the galaxies. 
To make predictions for these observables,  one has to know  the DM
density  in the region of our Galaxy where  the Sun is located as well as the velocity
distribution for DM particles.  The more impressive evidence of DM is the  flatness of  galactic rotation curves
at large distances $r$ from the galactic center
$$v(r)\approx Const$$
 which is established  from observations of a  large set of spiral galaxies.
See, for instance, Fig.\ref{fig:flat_curve}.

\begin{figure}
\begin{center}
\psfig{file=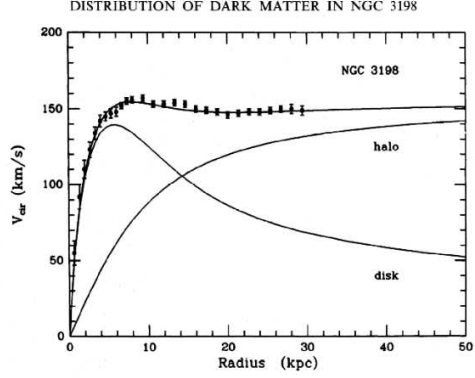,width=2.5in}
\end{center}  
\caption{ Rotation curve for spiral galaxy NGC 3198\cite{vanAlbada:1984js}.}
\label{fig:flat_curve}
\end{figure}

For the  rotation velocity to be constant at  all distances, assuming that
the DM contribution dominates, the density has to be  
\begin{equation}
\label{rho_r}  
   \rho(r)=\frac{v_{rot}^2}{4\pi G r^2}\;\;,
\end{equation}
where $G$ is the gravitational constant. DM velocity distribution  is related to
the spatial  distribution. In the simplest case
one can assume the micro-canonical DM phase space distribution\footnote{In general,
the equilibrium phase space distribution could also depend on angular momentum.}
$$  f(E_{tot}/M_{cdm}) d^3v d^3x= f( v^2/2 - v_{rot}^2 log(r)) d^3v
d^3x\;\;\;, $$
where $v_{rot}^2 log(r)$ is the gravitational potential of unit mass. This phase space distribution 
has to reproduce the DM density Eq.(\ref{rho_r})
$$ \int f( v^2/2 - v_{rot}^2 log(r)) d^3v = \frac{v_{rot}^2}{4\pi G r^2}$$
which leads to
$$ f(\mathcal{E}) = \frac{ exp(-2 \mathcal{E}/v_{rot}^2)}{ 4 \pi^{5/2} G
v_{rot}^2} $$
and  in particular to the Maxwell isothermal DM velocity
distribution
\begin{equation}
\label{Maxwell}
   \frac{e^{-(v/v_{rot})^2}d^3v} {\pi^{3/2}v_{rot}^3}    
\end{equation}

For remote galaxies, the  rotation curve is well measured via the Doppler shift  of the 21.1
cm line of hydrogen radiation. For the Milky Way, the  rotation curve is not known as
well as for other galaxies,   see Fig.\ref{fig:MW_curve}
  
\begin{figure}
\begin{center}
\psfig{file=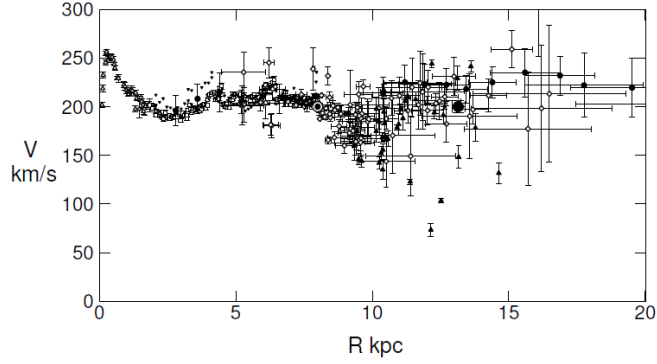,width=3.5in}
\end{center}
\caption{ Rotation curve for the Milky Way. Compilation of different
experiments presented in Ref\cite{Sofue:2008wt}  and normalized for
$R_\odot=8 {\rm kps}$, $v_{rot}(R_\odot=200 {\rm km/s}$.}
\label{fig:MW_curve}
\end{figure}

The Sun is   located  at a  distance $R_\odot=8.4kpc$  from the galactic center, 
where the rotation curve becomes flat, with \cite{Nakamura:2010zzi}
$$v_{rot} \approx 240(10)  {\rm km/s}\;\;.$$ 
For this value of the rotation velocity,  Eq.(\ref{rho_r}) gives an estimation $\rho\approx0.58 GeV/cm^3$. A more
precise estimation of the  DM density at the Sun orbit is    
 $$           \rho_{\odot} = 0.3 {\rm GeV/cm}^3 $$  
within a factor $2-3$ uncertainty\cite{Nakamura:2010zzi}. Because the Galaxy
has a finite gravitational potential, there is some escape velocity
corresponding to Sun orbit. Observations of fast  stars in the Milky Way give
the limits of \cite{Nakamura:2010zzi}
   $$   498{\rm km/s} <  v_{esc} <608 {\rm km/s}.$$
   The Maxwell velocity distribution is truncated at $v_{esc}$. 

micrOMEGAs has several  global parameters which describe the Dm distribution in the Galaxy, see Tab.\ref{tab:global}.
\begin{table}
\tbl{ Galaxy parameters}
{
\begin{tabular}{|l|c|c|c|l|}
\hline
Name  & default value & units &Symbol&Comment\\
\hline
Rsun  & 8.5           & kpc   &  $R_\odot$      &Distance from Sun to center of Galaxy\\
rhoDM & 0.3           & GeV/cm$^3$&$\rho_\odot$ &Dark Matter density at Rsun\\
Vesc  & 600           & km/s  &  $v_{esc}$      &galactic escape velocity        \\
Vearth&  225.2        & km/s  &  $v_{Earth}$    &Galaxy velocity of Earth     \\
Vrot  & 220           & km/s  &  $v_{rot}$      &rotation velocity at Sun orbit \\
\hline
\end{tabular}
}
\label{tab:global}
\begin{tabnote}
We keep the old recommended 1985 IAU value for $v_{rot}$ because it is 
used for  comparing results of Direct Detection experiments\\
\end{tabnote}
\end{table}

In general different DM  spatial and velocity distributions in the Milky Way are
considered and micrOMEGAs can work with any distribution implemented as
external functions. But the formulas   presented above
 characterize the  DM distribution and contain all important
features  of distributions used in real calculations of direct and indirect DM
signals. 

The DM  spatial mass  density  in micrOMEGAs  is  given as a product of the local density at the
Sun orbit with the halo profile function.
$$    \rho(r)=\rho_{\odot}F_{halo}(r) \;\;.$$ 
By default micrOMEGAs uses the Zhao\cite{Zhao:1995cp} profile  
\begin{equation}
\nonumber
F_{halo}(r)=\left(\frac{ R_{\odot}}{r}\right)^{\gamma}
\left(\frac{r_c^{\alpha}+ R_{\odot}^{\alpha}}
{r_c^{\alpha}+r^{\alpha}}\right)^{\frac{\beta -\gamma}{\alpha}}
\end{equation}
with parameters 
$\alpha=1,\beta=3,\gamma=1,rc=20[kpc]$ which corresponds to  the NFW\cite{Navarro:1996gj} profile.
 These values can be changed by
the command \\
\verb|     setProfileZhao(|$\alpha$,$\beta$,$\gamma$,$r_c$\verb|)| ,\\
 for example, \\
\verb|     setProfileZhao(2,2,0,3.5)| \\
 sets isothermal profile with a core\cite{Bergstrom:1997fj}.
micrOMEGAs can work with any  external function that describes a spherically symmetric  halo profile. 
For this,  call \\
\verb|     setHaloProfile|($myHaloProfile$)\\
with  the name of external functions as argument. To restore the default
Zhao profile  use \\
\verb|     setHaloProfiles(hProfileZhao)|.

 \section{Direct Detection in micrOMEGAs.}

\subsection{ Amplitudes in the $v=0$ limit.}
To predict the  Direct Detection rate one 
has to  calculate the differential cross section for elastic
scattering  of a DM particle  on atomic nuclei. 
The velocities of  DM particles near the Earth are close to the
orbital velocity of the Sun, $v\approx 0.001c$.
Since elastic cross sections are finite in the  $v\to 0$ limit, we can 
compute the DM nucleon cross sections in this limit thus simplifying the computation.
However the  transfer momenta are large as compared 
with the inverse size of a nucleus, so nuclei elastic form factors have to be taken
into account.  In the non-relativistic  limit the DM-nucleon elastic amplitudes
can be divided into two classes, the scalar  or spin independent (SI)
interaction and the axial-vector or spin dependent (SD) interaction. 
For a spin 1/2 nucleon interactions corresponding to higher spin exchange 
will clearly vanish in the zero momentum limit. When the DM particle is not
self-conjugate  the amplitudes can be further divided in two classes 
{\it even} and {\it odd} with respect to swapping $ DM \leftrightarrow \overline{DM}$.

DM-nucleon amplitudes are related to the 
DM-quark amplitudes after introducing form factors that describe the quark content of the nucleon.
These form factors are different for different quark current.
The general scheme for  calculating  DM nuclei cross sections is the following:
\begin{romanlist}
\item we expand the DM-quark amplitudes over basic operators  in  the $v=0$ limit;
\item using information about nucleon quark form factors we transform 
DM-quark amplitudes into  DM-nucleon amplitudes;
\item   we use the nucleon form factors in nuclei to  calculate the DM-nuclei 
differential  cross sections. 
\end{romanlist}

\subsection{DM-quark operators.}
\label{operators}
The independent set of DM-quark operators  in the $v=0$ limit are presented in
Tab.\ref{tbl:DDoperators}, 
\begin{table}
\tbl{ Operators for DM($\chi$) - quark  interactions.}
{\begin{tabular}{|c|l|c|c|}
\hline
&DM &    $\hat{\mathcal O}_{e}$  & $\hat{\mathcal O}_{o}$ \\
&Spin&  Even operators      &  Odd operators   \\
\hline
&&&\\ 
& 0  & $2M_\chi\phi_\chi \phi_\chi^* \overline{\psi}_q\psi_q $ & $
i (\partial_{\mu} \phi_\chi \phi_\chi^* - \phi_\chi
\partial_{\mu}\phi_\chi^*)\overline{\psi}_q\gamma^\mu \psi_q $ \\
 SI &1/2 & $\overline{\psi_\chi}\psi_\chi  \overline{\psi}_q\psi_q $  &
$\overline{\psi}_\chi\gamma_\mu\psi_\chi
 \overline{\psi}_q\gamma^\mu\psi_q   $ \\
& 1  & $ 2 M_{\chi} A^*_{\chi\mu} A_\chi^{\mu}
  \overline{\psi}_q\psi_q$ & +$i \lambda_{q,o}(A_\chi^{*\alpha}\partial_\mu
A_{\chi,\alpha}
-A_\chi{^\alpha}\partial_\mu
- A_{\chi\alpha}^*)$\\
  & & & $\overline{\psi}_q\gamma_\mu\psi_q$      \\   
\hline\hline
&&&\\
&1/2 &$ \overline{\psi}_\chi\gamma_\mu\gamma_5\psi_\chi
\overline{\psi}_q\gamma_\mu\gamma_5\psi_q $
& $-\frac{1}{2}\overline{\psi}_\chi\sigma_{\mu\nu}\psi_\chi
\overline{\psi}_q\sigma^{\mu\nu}\psi_q $  \\
SD& 1  & $\sqrt{6}(\partial_\alpha A^*_{\chi\beta} A_{\chi\nu} -
A^*_{\chi\beta} \partial_\alpha A_{\chi\nu})$
  &
$i\frac{\sqrt{3}}{2}(A_{\chi\mu}  A_{\chi\nu}^* - A_{\chi\mu}^*
A_{\chi\nu} ) \overline{\psi}_q\sigma^{\mu\nu}\psi_q$
\\
&  &
$\epsilon^{\alpha\beta\nu\mu}\overline{\psi}_q\gamma_5\gamma_\mu\psi_q$  &\\
\hline
\end{tabular}
}
\label{tbl:DDoperators}
\end{table}

To compute the amplitudes for DM -quark scattering at low energy, micrOMEGAs expands the Lagrangian of the model
in terms of local operators and extracts the coefficients of the low energy effective DM quark Lagrangian numerically.
To do this  we need to add to the model Lagrangian the projection operators defined in Table 1. The interference between one projection operator and the effective vertices will single out either the spin dependent or spin independent contribution, since the effective Lagrangian is written in an orthogonal basis.
The matrix element 
\begin{equation}
\label{operatorEQ}
\langle q(p_1),\chi(p_2)| \hat{S} {\hat{\mathcal O}_{e,o}} |q(p_1),\chi(p_2) \rangle
\end{equation} is then evaluated numerically
at zero momentum transfer. Here  $\hat{S}$ is the S-matrix  obtained from the complete Lagrangian 
at the quark level.


In the general case, there are other operators with higher order derivatives
which contribute in the  $v=0$ limit, in particular {\it twist=2}  operators.
The contribution  of twist=2 operator is however suppressed by a factor
$\frac{M_N}{M_{\widetilde{q}} -M_{cdm}}$ where $M_N$ is the nucleon mass,
$M_{\widetilde{q}}$ is the mass of the lightest odd particle which couples to DM and quarks.
To  extract the contribution of  twist=2 operators, micrOMEGAs calculates the matrix
element (\ref{operatorEQ}) for non-zero collision momenta. 
 
\subsection{Nucleon form factors.}
Each operator describing  DM-quark interaction is related to  a similar operator for
DM-nucleon interactions with a coefficient called the {\it nucleon form factor}.
Some of the   form factors  of  light quarks    are  determined experimentally, other
can be obtained by  lattice calculations. micrOMEGAs uses global
parameters to store the form factors of light quarks thus allowing the user to 
change them easily.  The list of nucleon form factors is presented in Tab.\ref{tbl:NFF}.

\subsubsection{Even nucleon form factors}
Even form factors give the same contribution to $DM$ and $\overline{DM}$
scattering,  they include 
\begin{itemlist}

\item {\bf Scalar (SI) even form factor for light quarks.} 
These form factors,  $f_q^N$, which relate quark and nucleon operators  are defined by  
\begin{equation}
\label{Sqq}
\langle N|  m_q \overline{\psi}_q \psi_q|N \rangle = f_q^N M_N 
\end{equation}
where N means nucleon and $M_N$  its mass.
 The scalar form factor 
actually counts  the contribution  of each quark  to the  nucleon mass.   If, for instance, the DM
interacts with light quarks via Higgs exchange, then the matrix element  will  contain a 
quark mass. This mass will cancel with the quark mass in  the  scalar form factor so that the final answer will be independent of the quark mass, a quantity  which depends strongly on the QCD scale. This also means that
despite the  very small coupling of light quarks with the Higgs,  they give a  finite contribution to 
DM-nucleon scattering.

Scalar form factors are known from hadron spectroscopy, data of $\pi N$
scattering\cite{Pavan:2001wz} and lattice calculations of {\it s} quark contribution to the 
nucleon mass. Still there is some uncertainty  in the form factors especially  for the  s-quark contribution.

\item {\bf Axial-vector(SD) even  form factors.}
The axial-vector current $\overline{\psi}_q\gamma_{\mu}\gamma_5\psi_q$ counts
the total spin of quarks and anti-quarks {\it q} in the nucleon.  
The axial-vector interactions in the nucleon are related to those involving quarks, with
\begin{equation}
2s_\mu \Delta q^N=\langle N|\overline{\psi}_q\gamma_{\mu}\gamma_5\psi_q|N \rangle
\end{equation}
where $s_\mu$ is the nucleon spin.  The coefficients $\Delta q^N$ are extracted from 
   HERMES\cite{Airapetian:2007mh} and COMPASS\cite{Ageev:2007du} experiments, they are defined  as \verb|pVectorFFPq| in micrOMEGAs, 
see Tab.\ref{tbl:NFF}.

\item {\bf Form factors for heavy quarks.}

The nucleon consists of light quarks and gluons, nevertheless one can also consider interactions of DM with heavy quarks inside the nucleon. These effectively come into play since heavy quark loops contribute to the interactions of DM with gluons, see for example the diagrams in Fig. \ref{fig:QuarkLoops}. Heavy quarks, Q, give a large contribution to the SI amplitude. Their form factor can be calculated in QCD.

\begin{figure} 
\begin{center}

\includegraphics[height=4.cm, width=8cm,angle=0]{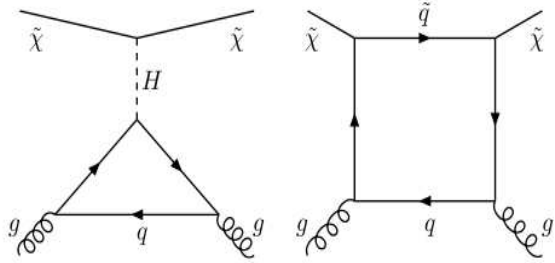}
\caption{Diagrams that contribute to DM-gluon
interaction via heavy quark loops } 
\label{fig:QuarkLoops}
\end{center}
\end{figure}

When the contribution of the triangle diagram in Fig.\ref{fig:QuarkLoops} is dominant
one can use the anomaly of the trace
of the QCD energy-momentum tensor\cite{Shifman:1978zn} to calculate the heavy quark 
form factor. The leading order result is 
\begin{equation}
 \label{shifman}
 f_Q=\frac{2}{27}
\end{equation}
which is about twice as large than the form factor of light quarks. If the mass of $\widetilde{q}$, the new particle  that couples to the DM and a quark,  is not large then this simple formula cannot be used and  one has to include the contribution
of  box diagrams. The box diagrams in the MSSM were computed by Drees\&Nojiri\cite{Drees:1993bu}. Because QCD interaction is universal 
these formulas   can be used in any model with spin 1/2 DM and spin
0 $\widetilde{q}$.

The contribution of heavy  quarks to SD amplitudes is  negligible. 
\end{itemlist}

\begin{table}
\tbl{ Nucleon form factors.}
{
\begin{tabular}{|c|clc|cl}
\hline
Quark& \multicolumn{2}{|c|}{Even}&\multicolumn{1}{|c|}{Odd} \\
\hline
&SI&SD&SD\\
\hline
\multicolumn{4}{|c|}{Proton}\\
\hline
&  ScalarFFP{\it q}  &   pVectorFFP{\it q} &   SigmaFFP{\it q}  \\
\hline
d & 0.0253   &   -0.427 &   -0.23 \\
u &   0.0205   &    0.842 &  0.84 \\
s &  0.0277    &   -0.085 &    -0.046\\
\hline
\multicolumn{4}{|c|}{Neutron}\\
\hline
&  ScalarFFN{\it q}  &   pVectorFFN{\it q} &   SigmaFFN{\it q}  \\
\hline
d &  0.0370   &  0.842  &    0.84   \\
u &  0.0140   &  -0.427 &    -0.23   \\
s &   0.0277    &   -0.085 &    -0.046  \\
\hline
\end{tabular}
}
\begin{tabnote}
 Names of global parameters for  form
factor can be obtained by replacing $q$ with the  quark
name.\\
\end{tabnote}\label{tbl:NFF}

\end{table}

\subsubsection{Odd nucleon form factors}

Odd form factors are needed to calculate the  difference between $DM-N$  and
$\overline{DM}-N$ amplitudes.  
\begin{itemlist}

\item {\bf Scalar (SI) odd form factors}. The $\overline{\psi_q}\gamma_{\mu}\psi_q$ quark 
current gives rise to scalar interactions. Since the vector current is conserved,  the associated  nucleon form factor just
counts the number of quark minus the number of antiquarks in the nucleon. Thus the  \{{\it u },{\it d}\}
quarks have form factors \{2,1\} in proton and \{1,2\} in neutron while the form factor is zero for heavy quarks .  

\item {\bf Pseudo-vector (SD) odd form factors.} 
 The $\langle N|\overline{\psi} \sigma_{\mu\nu}\psi|N\rangle$  current can be
interpreted as the difference between the spin of quarks and the spin of anti-quarks in
nucleons. Measurements by COMPASS  and HERMES indicate that the antiquark
contribution to nucleon spin is  compatible with zero. Thus these form
factors should be close to the even SD axial-vector form factors.  QCD  lattice
calculations\cite{Aoki:1996pi,Dolgov:2002zm} confirm this expectation, the lattice results are used as default values 
for the form factors  \verb|SigmaFFPq|  listed in 
Tab.\ref{tbl:NFF}. 

\end{itemlist}

{\bf Form factors for twist=2 operators} are calculated via parton 
distribution functions 
$$ f_q^{twist=2} = \frac{1}{2} \int_0^1 dx x f^{pdf}_q(x,Q=M_q)$$

\subsection{Nucleon amplitudes: example.}
\label{DMnucleonExampe}
In micrOMEGAs the amplitudes for DM-nucleon scattering   at rest can be computed using the
routine \\
\verb|     nucleonAmplitudes(qBOX,pAsi,pAsd,nAsi,nAsd)|\\
where {\it pAsi, pAsd, nAsi, nAsd} are output parameters which contain
proton/neutron  SI/SD amplitudes. Each of these parameters is a two 
dimensional array which contains 
amplitudes for $DM$ and $\overline{DM}$ scattering.  Amplitudes are normalized
so that the total cross section for $DM$-nucleon scattering reads 
$$ \sigma_{tot}=\frac{4M_{cdm}^2 M_N^2}{\pi(M_{cdm}+M_N)^2}(|pAsi[0]|^2+3|pAsd[0]|^2)$$

The parameter  \verb|qBOX| specifies the name of the
 function which calculates the box diagram of Fig.\ref{fig:QuarkLoops}. 
The  option \verb|qBOX=FeScLoop|  includes the calculation of the  
 box diagram with a   spin 1/2 DM and a scalar
$\widetilde{q}$ particle. When \verb|qBOX=NULL|, 
the  form factors for heavy quarks (\ref{fig:QuarkLoops}) with QCD  NLO
correction are  used. This approximation works well when the masses of $\widetilde{q}$  particles  
 included in vertices $ DM\cdot q \cdot \widetilde{q}$  are large. \\

For the MSSM test point of section \ref{MSSM_example} we have 
{\footnotesize
\begin{verbatim}
//            CODE                                        OUTPUT
nucleonAmplitudes(FeScLoop, pA0,pA5,nA0,nA5);       |
printf("CDM-nucleon micrOMEGAs amplitudes:\n");     |CDM-nucleon micrOMEGAs amplitudes:
printf("proton: SI=%.2E SD=%.2E\n",pA0[0],pA5[0]);  |proton: SI=-1.33E-09 SD=-1.58E-08
printf("neutron:SI=%.2E SD=%.2E\n",nA0[0],nA5[0]);  |neutron:SI=-1.33E-09 SD= 1.98E-08 
                                                    |
coef=4/M_PI*3.8938E8*pow(Nmass*Mcdm/(Nmass+Mcdm),2);|
printf("CDM-nucleon cross sections[pb]:\n");        |CDM-nucleon cross sections[pb]: 
printf(" proton  SI %.3E  SD%.3E\n",                |
    coeff*pA0[0]*pA0[0],3*SCcoeff*pA5[0]*pA5[0]);   |proton  SI 2.177E-10  SD 3.223E-07
printf(" neutron SI %.3E  SD %.3E\n",               |
    coeff*nA0[0]*nA0[0],3*SCcoeff*nA5[0]*nA5[0]);   |neutron SI 2.696E-10  SD 5.113E-07
\end{verbatim}
} 
[A
The cross section for SI interactions on protons in this example, 
$$\sigma^{SI} \approx 2.2\cdot 10^{-10}[pb]= 2.2\cdot10^{-46}[cm^2]$$   
is well below the current Xenon100\cite{Aprile:2011hi} upper limit  $ \sigma^{SI} < 2\cdot 10^{-44}[cm^2]$,
but within reach of the next generation of  direct
detection experiments such as  Xenon1T.  

\subsection{Dark matter nucleus scattering.}
For zero DM velocity, the  cross section for DM-nucleus SI interaction reads 
\begin{equation}
\label{SI_at_rest} \sigma_0^{SI} =\frac{4\mu^2}{\pi} 
\left(\lambda_p Z + \lambda_n(A-Z)\right)^2\;,\;\;\;
\mu=\frac{M_{cdm}M_A}{M_{cdm}+M_A}
\end{equation}
where $\lambda_p$, $\lambda_n$ are amplitudes for DM scattering on nucleons;
$M_A$, Z, A are  the nucleus mass, charge, and atomic number respectively.
For a small DM velocity, $v\approx 10^{-3} c$, we neglect the dependence on the small momentum transfer in the cross section but include this dependence in the nucleus form factor, the differential cross section is
\begin{equation}
  \frac{d\sigma^{SI}}{dE} =
\frac{\sigma_0^{SI}}{E_{max}}F_A^2(q)\;,\;\;0<E<E_{max}=2\left(\frac{v^2\mu^2}{M_A}\right)
\label{eq:si}
\end{equation}
where $E$ is the nucleus recoil energy and $q=\sqrt{2EM_A}$ the transfer momentum.
The form factors $F_A(q)$ are well known from experiments of  $\mu$ scattering
on atomic nuclei. Note that Eq.~\ref{eq:si} predicts an  $A^2$ enhancement  of the SI cross
section at large A.  Such enhancement does not occur for  SD interactions  due to a 
 strong compensation of the  proton/neutron spins   with the same orbital
state.

 For SD  scattering  on nucleus,  three form factors are introduced
\begin{equation}
\label{SD_nucleus}
 \frac{d\sigma^{SD}_A}{dE}=\frac{16\pi\mu^2}{(2 J_A + 1)E_{max}}
( S_{00}(q)a_0^2 + S_{01}(q)a_0a_1 
  -S_{11}(q)a_1^2)  
\end{equation}
where $a_0= \xi_p+\xi_n$, $a_1= \xi_p-\xi_n$,   $\xi_p$, $\xi_n$ are proton and neutron SD amplitudes, $J_A$ is a total
angular momentum of nucleus. 
The $S_{ij}$ form factors are obtained by computation. The list of available
form factors is presented  in the review\cite{Bednyakov:2004xq,Bednyakov:2006ux}.  

\subsection{Calculation of Direct Detection signal in micrOMEGAs.}

The  distribution of the number of events over the recoil energy is obtained after integrating 
Eq.~(\ref{eq:si}, \ref{SD_nucleus})
over the  DM velocity distribution. In the resulting formulas, the DM local density enters  as a total
factor. In the micrOMEGAs package this density is a 
global parameter \verb|rhoDM|  whose default value is $0.3 GeV/cm^3$.
The recoil energy distribution is calculated by the routine\\ 
\verb|     nucleusRecoil(f,A,Z,J,Sxx,qBOX,dNdE)|\\
Here the input parameters are 
\begin{itemlist}
\item[$\diamond$]
\verb|f| -  the DM velocity distribution   normalized such that 
\begin{equation}
\label{velocity_norm}
 \int_0^{\infty} v f(v) dv =1
\end{equation}
The units  are $km/s$ for v and $s^2/km^2$ for  f(v).
\item[$\diamond$]
\verb|A| - atomic number of nucleus;
\item[$\diamond$]
\verb|Z| - number of protons in the nucleus;
\item[$\diamond$]
\verb|J| - nucleus spin;
\item[$\diamond$]
\verb|Sxx| - routine which calculates SD nucleus form factors,
Eq.(\ref{SD_nucleus}), for
spin-dependent interactions.
\item[$\diamond$]
\verb|qBOX| - a parameter needed by \verb|nucleonAmplitudes|, see the description above.
\end{itemlist}
The micrOMEGAs package contains  the predefined constants $Z\_\{Name\}$, 
$J\_\{Name\}\{atomic\_number\}$ for the electric charges and 
spins for a large set of isotopes, for instance, \verb|Z_Xe|, \verb|J_Xe129|,
\verb|J_Xe131|. Furthermore all SD form factors presented in the 
Review\cite{Bednyakov:2004xq,Bednyakov:2006ux} are included. Their generic name  is  $Sxx\_\{Name\}\{atomic\_number\}$, for
instance \verb|SxxXe129|. The isotopes whose charges, spins and 
form factors  are included in micrOMEGAs are\\
{\small
\verb|F19 Na23 Na23A Al27 Si29 K39 Ge73 Ge73A Nb92 Te125 I127 Xe129 Xe131|\\ 
}
For the velocity distribution, the user can substitute  any  function. The default function included in 
  micrOMEGAs is 
$${\rm Maxwell}(\mbox{v})=
\frac{c_{\rm{norm}}}{\mbox{v}}\int\limits_{|\vec{v}|<v_{esc}} d^3\vec{v} \exp\left(
-\frac{(\vec{v}-V_{Earth})^2}{v_{rot}^2}\right)\delta(\mbox{v} -|\vec{v}|)
$$
which  corresponds to the velocity distribution of the isothermal model.
Here  $V_{Earth}$, $v_{rot}$, and $v_{esc}$  a  global parameter presented
in Tab.\ref{tab:global}.

\verb|nucleusRecoil| returns the total  number of events per day and per kilogram of 
detector material. In general the result is averaged over 
$DM$ and $\overline{DM}$. The energy spectrum of recoil nuclei is stored in 
\verb|dNdE| array which contains 200 elements. The value in the $i^{th}$ element corresponds to
$$
dNdE[i] = \frac{dN}{dE}|_{E=i*keV}
$$
in units of ${\rm (1/keV/kg/day)}$. The recoil energy distribution can be 
displayed on the screen with \\
 \verb|     displayRecoilPlot(dNdE,title,E1,E2)|\\
where \verb|title|
is a character string specifying the title of the plot and
\verb|E1,E2| are minimal and maximal values for the displayed
energy in keV.  The function \\
\verb|     cutRecoilResult(dNdE,E1,E2)|
calculates the number of events in an energy interval defined by the
values \verb|E1,E2| in keV, each experiment normally gives an energy interval for their result.

\subsection{Example.}
For the  MSSM test point of  section \ref{MSSM_example}, to obtain the recoil energy corresponding to the Xenon100
experiment, we call 
\begin{verbatim}
double E1=8.4 /*KeV*/, E2=44.6/*KeV*/, Exposure=1471/*day*kg*/;
double nEvents, dNdE[200]; /* output */
int i;
nucleusRecoil(Maxwell,131,Z_Xe,J_Xe131,SxxXe131,FeScLoop,dNdE);
nEvent=Exposur*cutRecoilResult(dNdE,E1,E2);
printf("Expected number of events %.2E\n",nEvents);
for(i=0;i<200;i++) dNdE[i]*=Exposure;
displayRecoilPlot(dNdE,"Recoil energy distribution for Xe",E1,E2);
\end{verbatim}

The number of events given in output is $0.07$, this means that this point can be tested by XENON if one increases the exposure by roughly a factor
50, in agreement with the  estimate  obtained  in section \ref{DMnucleonExampe}. 
The resulting distribution  is displayed in  Fig.\ref{fig:XePlot}. Here we see that the number of events 
decrease rapidly with  the recoil energy. This is due to the  nucleus form
factor and to the DM velocity distribution.

\begin{figure}
\begin{center}
 \includegraphics[height=6cm, width=10cm,angle=0]{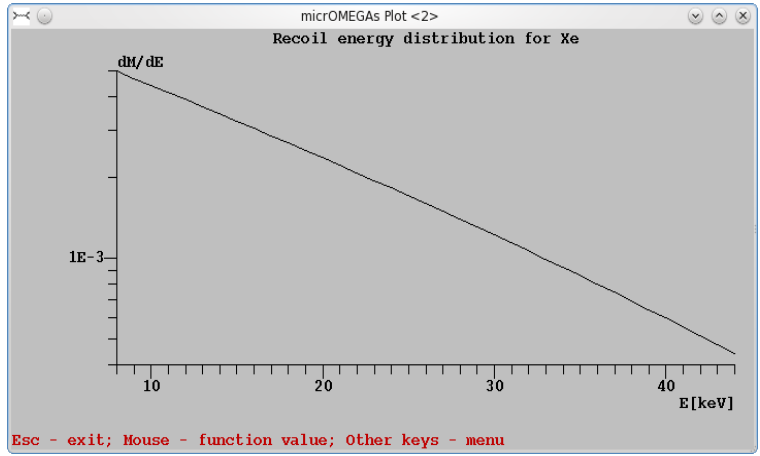}
 \vspace{.2cm}
 \caption{Recoil energy distribution for neutralino-Xenon collisions.}
\label{fig:XePlot}
\end{center}
\end{figure}

\subsection{Nuclei event rates for arbitrary DM-nucleon cross-sections.} 

The function\\
\verb|     nucleusRecoilAux(f,A,Z,J,Sxx,csIp,csIn,csDp,csDn,dNdE)|\\
is similar to \verb|nucleusRecoil|  except that \verb|csIp(csIn)|,
the  SI cross
sections for DM  scattering on protons(neutrons) and
\verb|csDp(csDn)|, the SD cross sections on protons(neutrons) are specified as
 input parameters (in {\tt pb} units). A negative value for one of these cross sections is interpreted as a
destructive interference between the
proton and the neutron amplitudes. Note that the rate of recoil energy  depends
implicitly on the {\tt Mcdm}  and {\tt rhoDM} parameters. 

The \verb|etc/| subdirectory contains the {\it DD\_ALL.c } program which can be 
compiled by the standard procedure \\
\verb|     make main=DD_ALL.c|\\
giving the executable {\tt DD\_ALL} which needs the
following command line arguments \\
\verb|     ./DD_ALL   Mcdm  csSIp  csSIn  csSDp csSDn |\\
This call  writes the prediction  for DM detection in DAMA\_NAI, Xenon, CDMS,
CoGent and COUPP. Some experimental input data used in this program 
(exposures, energy cuts, efficiency) can be obsolete. The user 
can improve this code easily to take into account new experimental data.

\section{Neutrino signals of  DM captured by Sun and Earth.}
\subsection{Neutrino fluxes}
Dark Matter  particles captured by the Sun/Earth accumulate in the center of
the Sun/Earth and annihilate into standard model particles. The neutrinos that escape the Sun/Earth can then be detected. When the capture and annihilation rates reach an equilibrium (we assume that equilibrium can be reached  within the lifetime of the Sun/Earth),
the  rate of DM annihilation is defined by the capture rate. This rate depends on the
cross sections for elastic scattering of DM on nuclei in the Sun/Earth. The
larger the atomic weight, the larger the   DM  energy loss. Knowing the 
chemical contents of the Sun/Earth is important  for the calculation of the capture rate.
The $DM$($\overline{DM}$) capture rates $C_{\chi}$
($C_{\bar{\chi}}$) depend also on  the DM density and
velocity distribution at the Sun galactic orbit. 
To calculate the capture rate  we
use  the simplified  formula found in the review of  Jungman {\it et.al.}\cite{Jungman:1995df}. 

\begin{eqnarray*}
C_{\chi} &=& 4.8\times 10^{28}\,{\rm s}^{-1}  
               \left(\frac{\rho_{\chi}}{0.3\,{\rm GeV/cm^3}}\right) 
               \left(\frac{270\, {\rm km/s}}{\bar{v}}\right) \nonumber \\
   & & \times  \displaystyle\sum_{i} \left( \frac{\sigma_{{\chi} i}}{pb}\right) 
                 \frac{ f_i \phi_i}{m_{\chi} m_{N_i}} F_i(m_{\chi}) S(m_{\chi}/m_{N_i})
\end{eqnarray*}
where  $m_{N_i}$, the mass of the nuclear specie $i$, and $m_{\chi}$ are given in GeV. 
$\bar{v}$ is the DM velocity dispersion, $f_i$ is the mass fraction of element $i$ in the Sun, and $\phi_i$ its distribution. 
$F_i(m)$ is a form factor suppression and $S$ a kinetic suppression factor. 

Let $\Gamma_{\chi\chi}$,
$\Gamma_{\bar{\chi}\bar{\chi}}$ and $\Gamma_{\chi\bar{\chi}}$ be the  annihilation rates for 
$DM(\overline{DM})$  processes  in the Sun/Earth and $N_{\chi}/N_{\bar{\chi}}$ the $DM/\overline{DM}$ densities in the Sun/Earth. For a self-conjugate DM,
$$C_\chi= 2\Gamma_{\chi\chi},$$
while in general we  have the following equations 
\begin{eqnarray} 
\nonumber
C_\chi&=& 2\Gamma_{\chi\chi}+\Gamma_{\chi\bar{\chi}}\\
\nonumber
C_{\bar{\chi}}&=& 2\Gamma_{\bar{\chi}\bar{\chi}}+\Gamma_{\chi\bar{\chi}}\\
\nonumber
\frac{\Gamma_{\chi\chi}}{\Gamma_{\bar{\chi}\bar{\chi}}} &=&
\frac{\langle\sigma_v^{\chi\chi}\rangle}
{\langle\sigma_v^{\bar{\chi}\bar{\chi}}\rangle}
\left(\frac{N_{\chi}}
{N_{\bar{\chi}}}\right)^2\\
\nonumber
\frac{\Gamma_{\chi\chi}}{\Gamma_{{\chi}\bar{\chi}}} &=&
\frac{\langle\sigma_v^{\chi\chi}\rangle/2}
{\langle\sigma_v^{{\chi}\bar{\chi}}\rangle}
\left(\frac{N_{\chi}}
{N_{\bar{\chi}}}\right)
\end{eqnarray}
which can be solved to obtain the annihilation rates  $\Gamma_{\chi\chi}$, $\Gamma_{\bar{\chi}\bar{\chi}}$ ,
$\Gamma_{\chi\bar{\chi}}$.  

The neutrino spectrum that result from DM annihilation inside the Sun is different from annihilation in the vacuum 
 because some long-lived  particles can 
 interact with the Sun medium before  decaying. Furthermore the neutrino spectrum 
is deformed by 
 attenuation and  oscillation processes that occur during 
propagation inside the Sun/Earth.  In
Ref\cite{Cirelli:2005gh} all these factors  where taken into account and the 
neutrino spectra  after propagation  were tabulated for different DM masses. To get the
neutrino flux  at the surface of the Earth micrOMEGAs calculates for each SM final state
the relative contribution of the $\chi\chi\to SM SM$ channel and multiplies this by
the  annihilation rate $\Gamma_{\chi\chi}$ and the tabulated neutrino spectra  functions\cite{Cirelli:2005gh}
prepared for this channel for the relevant $M_{cdm}$.\\
\verb|     neutrinoFlux(forSun,nu_flux, Nu_flux)|\\
calculates $\nu_{\mu}$ and $\nu_{\bar{\mu}}$  fluxes  close to the Earth surface.  
If {\tt forSun==0} then the flux  of neutrinos from the Earth is calculated, otherwise 
this function computes the flux of  neutrinos from the Sun. The velocity distribution is assumed to be
Maxwellian and its parameters can be changed by \verb|SetfMaxwell| function.
The calculated fluxes  are stored in \verb|nu_flux|, \verb|Nu_flux| arrays of dimension
NZ=250. For  neutrino fluxes we use units \verb|[1/Year/|$km^2$\verb|]|.

\subsection{micrOMEGAs tools  to work with particle spectra.}
\label{toos_for_spectra}
Various  particle  spectra  and fluxes relevant to neutrino signals and to  indirect detection
ones (see next section)  are stored in arrays with {\bf NZ=250} elements.
The $i^{th}$ element of an array corresponds to 
 $dN/dz_i$ where $z_i=log(E_i/Mcdm)$\footnote{ The $z_i$ grid can be obtained with the function $Zi(i)$.}.
Spectra interpolation can be done by one of the two functions\\ 
\verb|     zInterp(z,flux)| which returns \verb|d(flux)/dz| and\\
\verb|     SpectdNdE(E,flux)| which returns \verb|d(flux)/dE[GeV]|\\
where \verb|flux| means the array where tabulated data are stored.
To display the  function  on the screen one can use \\
\verb|     displaySpectrum(flux,title,Emin,Emax,Units)| ,\\
where \verb|title| is a text string which gives a title to the  
plot. \verb|Emin| and \verb|Emax| define plot limits. If \verb|Units=0| the
flux is displayed as a function of  
$z_{10}=log_{10}(E/{\rm {Mcdm}})$ otherwise one gets the $dN/dE$ plot as a function of the energy
{\tt E} in GeV units.\\

The integrated  spectra/fluxes can be obtained by\\  
\verb|     spectrInfo(Xmin,flux,&Ntot,&Xtot)|
where $X_{min}=E_{min}/M_{cdm}$ and
\begin{eqnarray}
\nonumber 
 N_{tot} &=& \int_{X_{min}}^1 \frac{d(flux)}{dx}dx\\ 
\nonumber
 X_{tot} &=& \int_{X_{min}}^1 x\frac{d(flux)}{dx}dx
\end{eqnarray}.

\subsection{Muon fluxes}

A detector measures the muon flux that results from interactions of neutrinos with rocks below the detector or water within the detector. 
The routines which calculate such $\mu$ fluxes  are based on the formulas of  Ref.\cite{Erkoca:2009by}
\noindent
\verb|     muonUpward(nu_flux,Nu_flux, mu)|\\
calculates the muon flux which result from
interactions of neutrinos with rocks. Here  \verb|nu_flux| and \verb|Nu_flux| are input parameters which designate the
neutrino/anti-neutrino fluxes  calculated by {\tt neutrinoFlux}. 
 {\tt mu} is an array which stores  the sum of $\mu^+$ and $\mu^-$  fluxes. 
\noindent 
\verb|     muonContained(nu,Nu,rho, mu)|
calculates  the flux of muon which  are produced in a given (detector) volume.
It has the same parameters as the function  for the
calculation of upward events. {\tt rho} is the density of detector material
in $g/cm^3$ units. For instance, {\tt rho=1}   for water. The  array {\tt mu} gives the  muon
flux  in {\tt  1/Year/}$km^3$ units representing  the number of neutrinos
converted to muons in a $km^3$ volume. 


\subsection{Example}

As an example we calculate the muon flux for the CMSSM   with parameters\\
$$   M_0=3200,\; M_{1/2}=315,\;   A_0=0,\; \tan\beta=10, \;sign(\mu)=1,\; M_{top}=173.2;$$ 
The commands 
\begin{verbatim}
 double nu_flux[NZ],Nu_flux[NZ],mu_flux[NZ];
 double Ntot, Emin=1;
 neutrinoFlux(1/* for Sun*/,nu_flux, Nu_flux);
 neutrinoUpward(nu_flax,Nu_flux, mu_flux); 
 displaySpectrum(mu,"Upward muons[1/Year/km^2/GeV]",Emin,Mcdm/2,1);
 spectrInfo(Emin/Mcdm,mu, &Ntot,NULL);
 printf("Upward muon flux for   1GeV threshold %.2E /Year/km^2\n",Ntot);
 spectrInfo(50/*GeV*//Mcdm,mu, &Ntot,NULL);
 printf("Upward muon flux for  50GeV threshold %.2E /Year/km^2\n",Ntot);
\end{verbatim}
will generate the  muon spectrum displayed in Fig.\ref{fig:muFlux}.
and print integrated flux in \verb|1/Year/km^2| units
\begin{verbatim} 
Upward muon flux for   1GeV threshold=2.50E+03
Upward muon flux for  50GeV threshold=2.51E+02
\end{verbatim}
\begin{figure}
\begin{center}
 \includegraphics[height=6cm, width=10cm,angle=0]{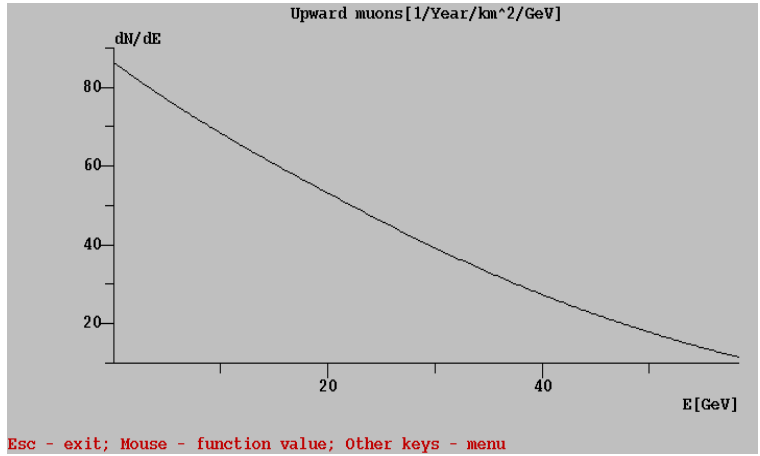}
 \vspace{-.3cm}
 \caption{ Solar muon flux.}
\label{fig:muFlux}
\end{center}
\end{figure}

\section{Indirect Detection in micrOMEGAs}

\subsection{Initial annihilation spectra.}
Should primordial self-annihilations take place in the early Universe, the same process would take place nowadays in the
denser regions of the Galactic DM halo. 
DM annihilation in the Galactic halo produces pairs of Standard Model particles that hadronize and decay into stable particles.
The final states with  $\gamma$, $e^+$, $\bar{p}$, and $\nu_{\mu}$ are particularly interesting as 
they are the  subject of indirect  searches.  The production rate of particles from DM annihilation at location
 {\bf x} reads
\begin{equation}
\label{eq:DMflux}
Q_a({\bf x},E)\;\;=\;\;\frac{1}{2} \langle\sigma v\rangle \frac{\langle \rho({\bf
x})^2\rangle}{M_{cdm}^2} f_a(E)\;\;,
\end{equation}
where $\langle\sigma v \rangle$ is the  annihilation cross-section 
times the relative velocity of incoming DM particles which we evaluate in the limit $v=0$ (this
is a good approximation since $v=10^{-3} c$). Note that $\langle\sigma v\rangle$ includes averaging over incoming
particles/antiparticles. Thus, for a Dirac DM  particle $\chi$ for which  $\sigma_{\chi\chi}=0$, $\langle\sigma
v\rangle=1/2\langle\sigma_{\chi\bar\chi}v\rangle$. 
$\rho({\bf x})$ is the DM mass density at the location ${\bf x}$ 
and  $f_a(E)=dN_a/dE$ is the energy distribution of the particle $a$ produced in one reaction.
micrOMEGAs first calculates the  cross sections for all basic processes with  two body final states.
 The tabulated fragmentation functions obtained by Pythia are then used to generate spectra for each final stable particle. 
The routine to compute  the spectra $f_a$ and $\langle\sigma v\rangle$ is \\
 \verb|     calcSpectrum(key,Sg,Se,Sp,Sne,Snm,Snl,&err)|\\                        
which returns   $\langle \sigma v \rangle$ in $cm^3/s$ units. The calculated spectra
for $\gamma$, $e^+$, $\bar{p}$, and neutrinos, 
are stored in arrays  \verb|Sg|, \verb|Se|, \verb|Sp|, \verb|Sne|, \verb|Snm|, \verb|Snl|
of dimension {\bf NZ}(See section \ref{toos_for_spectra}). Note that the spectra of particles and  anti-particles are identical. 
{\bf key}  is a switch, depending on its value micrOMEGAs 
\begin{itemlist}
\item[1] takes into account the polarization of W,Z bosons. It can be important
for MSSM-like models where W,Z are  transversely polarized;
\item[2] calculates  $2\to 2 +\gamma$ processes to improve spectrum of  high energy
photons. 
\item[4] prints on the screen the  calculated  cross sections.
\end{itemlist}
More than one option can be switched on simultaneously by adding the corresponding values for \verb|key|. 
A problem in the spectrum calculation will produce a non zero error code, $err\neq 0$. 

The annihilation processes 
$DM,{\overline{DM}} \to 2\gamma$ and $DM,{\overline{DM}} \to \gamma,Z$ are loop suppressed
but lead to a very distinctive signature, a monochromatic gamma-ray line.
micrOMEGAs can calculate the corresponding  cross section only for specific models,
for example the MSSM.  The routine to do this is \\
\verb|     loopGamma(&vcs_gz,&vcs_gg)| which calculates  $\langle \sigma v
\rangle$ for $\gamma\gamma$ and $\gamma Z$ processes  in $cm^3/s$ units.

%
%
%
%

\subsection{Propagation of  photons and neutrinos}
 
The flux of photons (neutrinos) can be evaluated as
\begin{equation}
   \Phi_\gamma(E,\phi) = \frac{\langle\sigma v \rangle}{M_{cdm}^2} f_\gamma(E) H(\phi)\;\; \label{flux}
\end{equation}
where the factor $H$ includes the integral  over the line of sight,
\begin{equation}
H(\phi) = \frac{1}{8\pi}\int_0^{\infty} dr
\overline{\rho^2}\left(\sqrt{r^2+r_\odot^2-2rr_\odot\cos\phi}\right)
\label{halo_photon}
\end{equation}
and $\phi$ is the angle in the direction of  observation. This flux can be
calculated by the routine \\ 
\verb|     gammaFluxTab(fi,dfi,sigmav,Sg,Sobs)|\\
where  \verb|fi| is the angle between the line of sight and the center of the
galaxy,   \verb|dfi| is half the cone angle which characterizes the detector resolution
(the solid angle is  $2\pi (1-cos({\rm dfi})$),  
 \verb|sigmav|  and \verb|Sg| are the annihilation cross  section  and the photon spectrum calculated previously
 by \verb|calcSpectrum|.
\verb|Sobs| is the flux  observed in [$(cm^2s)^{-1}$]units.

Note that  the neutrino spectra obtained by \verb|calcSpectrum| do not take into account
 neutrino oscillation. One expects the muon
neutrino  flux at the Earth 
$$ f_{\nu_\mu}^{Earth} = 0.22f_{\nu_e}^0 +  0.39(f_{\nu_\mu}^0
+f_{\nu_\tau}^0 ) \;\;,$$
where numerical  coefficients are defined by neutrino flavour
mixing\cite{Abbasi:2011eq} and $f_\nu^0$ fluxes are calculated by
\verb|calcSpectrum|. 

\subsection{Propagation of charged particles}

The propagation of charged particles depends on the structure of the galactic magnetic
field and is characterized by the diffusion equations.   We assume that the region of diffusion of cosmic rays is 
represented by a thick disk 
of thickness $2L$ and radius $R\approx20$~kpc, with a thin galactic disk in the middle of the thick disk.
The  parameters entering this 
equation are not well known, but they are constrained by observations of the   B/C ratio in  cosmic rays\cite{Maurin:2001sj}. 
Table \ref{tbl:diff} lists the default values for the propagation parameters,
together with  the MIN/MAX variations which are compatible with the  B/C observations.

\begin{table}
\tbl{Global parameters of micrOMEGAs responsible for propagation of charged
particles.}
{
\begin{tabular}{|l|l|l|l|l|l|}
\hline
  Name      &MIN & MED&MAX    & Units &  Comments \\
            &    & Default&       &       &           \\ 
\hline
Rdisk       & 20 & 20 &20   &kpc   & Radius of the galactic  \\
            &    &          &       &       & diffusion disk, $R$ \\
K\_dif      &0.0016         & 0.0112&0.0765 & $\frac{kpc^2}{Myr}$ & Diffusion coefficient $K_0$\\
L\_dif      &1   & 4        & 15    & kpc   & Half height of the galactic   \\
            &    &          &       &       & diffusion zone $L$\\
Delta\_dif  &0.85& 0.7      & 0.46  &       &Slope of diffusion coefficient, $\delta$\\ 
Tau\_dif    & $10^{16}$& $10^{16}$& $10^{16}$&  s   &Positron energy loss, $\tau_E$\\
Vc\_dif     &13.5 & 12      & 5  &  km/s &Convective velocity of \\
            &     &         &       &       & galactic vind , $V_C$\\
\hline
\end{tabular}
}
\label{tbl:diff}

\end{table}

The energy spectrum of positrons is obtained by solving  the diffusion-loss
equation keeping only the two dominant contributions:  space diffusion and energy losses,
 \begin{equation}
\label{diff_positrons}
- {\bf \nabla}\cdot\left( K(E) {\bf \nabla} \psi_{e^+} \right)-
\frac{\partial}{\partial E} \left( b(E) \psi_{e^+}\right) =Q_{e^+}({\bf x},E)
\end{equation}
Here 
$$K(E)=K_0 \beta(E) \left({\cal R}/1\;{\rm GV}\right)^\delta$$
where $\beta$ is the particle velocity and ${\cal R}=p/q$ its rigidity.
The positron loss rate $b(E)$ is dominated by synchrotron radiation in the galactic magnetic field and inverse Compton scattering on
stellar light and CMB photons, 
\begin{equation}
b(E)=\frac{E^2}{E_0\tau_E}
\end{equation}
where $\tau_E=10^{16}$~s is the typical energy loss time. Subdominant contributions from diffusive reacceleration and
galactic convection were shown to have an impact only below a few GeV's and are not included in our
treatment. The function\\
\verb|     posiFluxTab(Emin,sigmav,Se,Sobs)|\\
computes the positron flux at the Earth. Here \verb|sigmav| and \verb|Se| are values obtained by 
\verb|calcSpectrum|.  \verb|Sobs| is the positron spectrum after propagation
in units (s ${\rm cm}^2 {\rm sr})^{-1}$).  \verb|Emin| is 
the energy cut to be defined by the user. Note that
a low value for \verb|Emin| increases the computation time.

The energy spectrum of antiprotons is obtained by solving the diffusion equation 
\begin{equation}
\label{aproton_eq}
\left[ -K(E)\nabla^2+V_c\frac{\partial}{\partial z}
+ 2(V_c  + h \Gamma_{tot}(E))\delta(z)\right]
\psi_{\bar{p}}(E,r,z)=Q_{\bar{p}}({\bf x},E)
\end{equation}
An important difference with the positron case is that energy loss of antiprotons is negligible, 
while the convective galactic wind  and annihilation of
anti-protons in the interstellar medium  should be taken  into account. The
function\\
\verb|     pbarFluxTab(Emin,sigmav, Sp,  Sobs)| 
computes the antiproton flux and works like \verb|posiFluxTab|.

The spectra calculated above are interstellar ones. The spectra measured
close to the Earth can be calculated by \\  
\verb|     solarModulation(Phi, mass, stellarTab, earthTab)|\\
Here \verb|Phi| is the
effective Fisk potential in MeV, \verb|mass| is the particle mass,
\verb|stellarTab| describes the interstellar flux, \verb|earthTab|
is the calculated particle flux at the Earth orbit.

\subsection{Example}

Recent  PAMELA\cite{Adriani:2010rc} results on the antiproton flux are in good
agreement with
background calculations\cite{Donato:2001ms}. It means that antiprotons 
are mainly produced  by collision of protons\footnote{Proton spectrum is about
four order of magnitudes exceeds the antiproton one.} with the interstellar gas.
micrOMEGAs has  a routine which provides the   interstellar  antiproton flux
tabulated in Ref\cite{Maurin:2006hy}\\                           
\verb|      pBarBackgroundTab(pBarBgTab)| .\\
In order to compare this flux with PAMELA,  one has first to apply the
\verb|solarModulation| routine. 

The antiproton flux resulting from DM annihilation in the MSSM is generally small unless one assumes  a large clump factor. 
Thus the MSSM generally agree with the PAMELA antiproton data. 
As an example of a model that leads to a large antiproton flux,   we consider the IDM
point presented in section \ref{sec:command}. 
 This point corresponds to $\Omega h^2 =0.102$
and gives a nucleon cross section $\sigma= 1.4\cdot 10^{-45}{\rm cm}^2 $ compatible with direct
detection experiments. This point also features a DM mass which very slightly exceeds the
half of  Higgs mass. 
Thus, the annihilation cross
section at rest ( in the galactic halo) which is enhanced by the exchange of a Higgs near resonance, is much larger than the annihilation  cross  section  at freeze-out temperatures, see
Fig.\ref{fig:vsigma_idm}. 
The resulting antiproton flux for two sets of propagation parameters is displayed in Fig.\ref{fig:antiprotons}. With the MED parameters, the antiproton flux is excluded by PAMELA results. 

\begin{figure}
\begin{center}
\vspace{-.8cm}
 \includegraphics[height=6.5cm, width=10cm,angle=0]{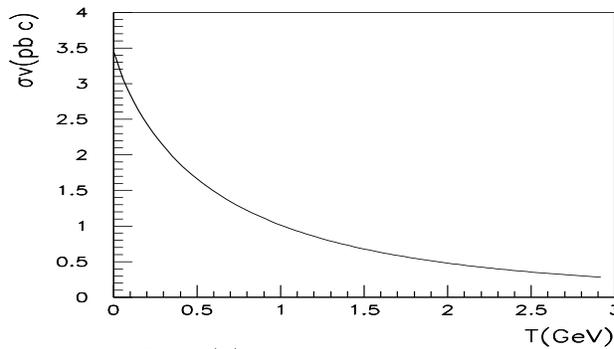}
 \vspace{-1.4cm}
 \caption{ $\sigma_v(T)$ for the sample IDM point of section \ref{sec:free_parameters}}
\label{fig:vsigma_idm}
\end{center}
\end{figure}

\begin{figure}
\begin{center}
 \includegraphics[height=7.2cm, width=10cm,angle=0]{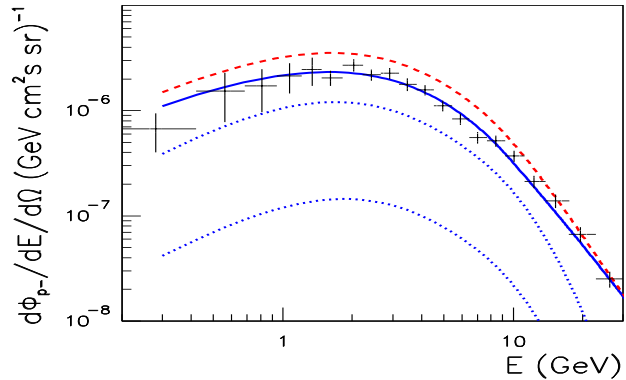}
 \vspace{-.8cm}
 \caption{ Differential flux of anti-protons in the IDM model point of section 3.5 assuming MIN (lower dot) and MED (upper dot) propagation parameters as compared to the background only (full line) and to the
measurement of PAMELA (crosses). The signal plus background for MED parameters is also displayed (dash).}
\label{fig:antiprotons}
\end{center}
\end{figure}

Below we present the list of micrOMEGAs commands which generate the data for the fluxes
displayed in  Fig.\ref{fig:antiprotons}
\begin{verbatim}
pBarBackgroundTab(BG);
solarModulation(560,1, BG,BG);
displaySpectrum(BG,"Background",Emin,Mcdm/2,1);
sigmaV=calcSpectrum(4,SpA,SpE,SpP,SpNe,SpNm,SpNl,&err);
pbarFluxTab(Emin, sigmaV, SpP,FluxP);
solarModulation(560,1, FluxP,FluxP);
displaySpectrum(FluxP,"antiproton flux(MED)" ,Emin,Mcdm/2,1);
K_dif=0.0016; L_dif=1; Delta_dif=0.85; Vc_dif=13.5;
  Rdisk=20; // MIN parameters 
pbarFluxTab(Emin, sigmaV, SpP,FluxP);
solarModulation(560,1, FluxP,FluxP);
displaySpectrum(FluxP,"antiproton flux(MIN",Emin,Mcdm/2,1);
\end{verbatim}

\section{Calculation of cross sections  in micrOMEGAs.}
The functions presented in this section are activated in micrOMEGAs template files  {\it
main.c[F]}  by uncommenting the instruction \\
\verb|     #define CROSS_SECTIONS  |\\
 micrOMEGAS uses the function\\ 
\verb|     numout*cc=newProcess(ProcessName);|\\
for generating codes of matrix elements for processes defined by the textual
variable {\it ProcessesName}. \verb|numout| is a special type to store
process code and auxiliary information. The variable \verb|cc| 
contains the computer address of the corresponding  code.   For instance\\
\verb|     cc = newProcess("e,E->2*x");|  \\
returns a code for the matrix elements of all processes with  $e^+e^-$ annihilation into two arbitrary
particles. {\tt newProcess} can generate the code for processes with more than two particles in the final state, however
the  current version of micrOMEGAs contains integration tools
only for $2\to 2$ processes. The function:\\  
\verb|     cs22(cc,L,Pcm,cos_min,cos_max,&err);| \\
calculates the cross section for $L^{th}$ subprocess where 
\begin{unnumlist}
\item {\tt Pcm} - momentum in center of Mass reference frame;
\item {\tt cos\_min}, {\tt cos\_max} - cuts for cosine of scattering angle in the same
frame;
\item {\tt L} numerates subprocesses. $1\le L \le ntot$,  the total number of
subprocess.
\end{unnumlist}
  The  number of the subprocess  as well as the numbers of incoming ($nin$) and outgoing ($nout$)
particles  for the compiled set of processes {\it cc}  can be obtained by \\
\verb|     procInfo1(cc,&ntot,&nin,&nout)|\\

The names of  incoming/outgoing particles of a given  process  and their  masses can be  obtained 
by the function\\
\verb|     procInfo2(cc,L,Names,Masses);|\\
where the variable {\tt char**Names} is an output parameter  used for  the  array of
particle names, and the variable {\tt double*Masses} is an array of numerical values for masses.

For proton-[anti]proton collisions we have to apply convolution of matrix
element with structure functions and perform summation over all types on
incoming partons.  This calculation can be performed with the
function\\
\verb|     hCollider( Pcm, pp, scale,  pName1, pName2)|\\
which  compiles the needed matrix elements and  calculates the cross section for 
production of two heavy particles at a hadron collider. Here {\it Pcm} is 
the beam energy  in the center-of-mass frame. If $pp>0$  then proton-
proton collision is considered, otherwise a  proton-antiproton collision. 
If \verb|scale=0| the largest mass of outgoing particles is used to set the QCD
scale. It is well known that  tree level   calculations with such scale  for 
{\it squarks} and {\it gluinos} production (in the MSSM) need a {\it
K-factor} about 1.5. If \verb|scale!=0| micrOMEGAs decreases the QCD scale to
simulate loop corrections.  
\verb|hCollider| uses the CTEQ6L set of parton distributions. The return value is the
cross section in \verb|[pb]| units.

For example, for the MSSM test point of section \ref{MSSM_example} a call of \\
\verb|     hCollider(3500.,1 ,1, "~uL","~UL" );|\\
gives the production cross section for  super-partners of  left-handed u-quark of 
0.0082pb at the LHC. The total cross section for all  channels which contain {\it squarks} and
{\it gluinos} gives 0.113pb. Note that this point has recently been excluded by SUSY searches in the jets plus missing $E_T$ channel by ATLAS and CMS.

\section{Calculation of particle widths in micrOMEGAs.}
The functions presented in this section are activated in micrOMEGAs template
{\it  main.c[F]} files with the instruction\\

\verb|#define DECAYS | \\
 Calculation of particle widths and decay branchings in  micrOMEGAs  can be
done by the functions \\
\verb|     pWidth( pName, &L)| where {\it pName} is name of particle and 
\verb|txtList L;|  is a variable of special type  intended to  store list of decay channels.\\
\verb|     findBr(txtList L, pattern)| allows to find the branching ratio of a specific channel. Here {\it pattern} is a text
variable which contains a list of decay particles separated by commas.      
If two particle decays are kinematically forbidden, micrOMEGAs
automatically calculates 3-body decays. 

The list of decay channels can be printed with the function \\
\verb|     printTxtList(txtList*L,FILE*f)|\\

Example:
{\small
\begin{verbatim}
              CODE                        |     OUTPUT
txtList L;                                |
double width;                             |       
width=pWidth("l",&L); // tau lepton       |       
printf("tau width=%.3E[GeV]\n",width);    |  tau width=2.052E-12[Gev]        
printf( "Branchings:\n");                 |  Branchings:                
printTxtList(L,stdout);                   |  5.998536E-01 l-> nl,U,d      
                                          |  2.000733E-01 l-> m,Nm,nl
                                          |  2.000733E-01 l-> e,Ne,nl
printf("Br(e,Ne,nl)=%E\n",findBr(L,       |
      "nl,Ne,e"));                        |  Br(e,Ne,nl)= 2.000733E-01           
\end{verbatim}
}

\section{ Implementation of new models in micrOMEGAs}

To implement new models in micrOMEGAs the user  has to 
\begin{romanlist}
\item create the model file structure presented in section \ref{file_structure} for the MSSM. 
  It can be done by one command launched in  micrOMEGAs root directory\\
   \verb|     ./newProject |{\it    myDarkMatter}\\ 
 It creates    the directory {\it myDarkMatter}   with all needed  subdirectories.   
\item  put CalcHEP model files into  subdirectory\\
\verb|     myDarkMatter/work/models/|.\\
 Recall that the    names of   odd sector particles have to  started with   the  \verb|\~|
symbol.
\item Launch \verb|./calchep| in \verb|myDarkMatter/| directory and use the 
{\it Check Models} menu line to test the model. In case CalcHEP returns an error message all mistakes 
should be corrected before starting the 
micrOMEGAs session.  
\item   If the  model constraints  or user's files {\it main.c}, {\it main.F}  need external functions, then the codes for these functions has to be 
put in the {\tt lib/} subdirectory.  The file \verb|lib/Makefile|  should  compile
them and create the  library {\it aLib.a}.   The default {\tt Makefile}  will  work properly
only if your functions are written in {\it C}. In general, the default {\tt Makefile}  has to be adapted. 

\item    Write function prototypes of {\it C} and/or {\it Fortran} functions in {\tt lib/pmodel.h}
and {\tt lib/pmodel.fh}  to use them  subsequently  in {\it main.c}  and {\it main.F}. 
\item Create an input file with the values of the free parameters of the model.

\end{romanlist}

General tools are available to make it easier to generate model files. The LanHEP~\cite{Semenov:2008jy}, 
FeynRules~\cite{Christensen:2008py} and SARAH~\cite{Staub:2008uz} packages allow automatic generation of model files in CalcHEP format. The former will be described briefly below as it was used to generate micrOMEGAs model files. 
Furthermore the SLHAplus package facilitates the implementation of complicated models.

\subsection{The  SLHAplus package.}
In several models of elementary particles we have noticeable loop
corrections to  particle masses. There are several packages which 
allow to perform such calculations for MSSM-like models: 
Isajet, SoftSusy\cite{Allanach:2001kg}, Spheno\cite{Porod:2003um}, SuSpect\cite{Djouadi:2002ze},
NMSSMTools\cite{Ellwanger:2006rn}~\footnote{ SuSpect and NMSSMTools packages
are included in micrOMEGAs} . There is an agreement to pass the calculated 
particles spectra and mixing matrices  via text file in a special format {\tt
SLHA}\cite{Skands:2003cj,Allanach:2008qq}. CalcHEP/micrOMEGAs contain the  package {\it
SLHAplus}\cite{Belanger:2010st}  intended   for 
reading SLHA files. An SLHA file also can contain information about
particle widths and decays channels. If such information is downloaded,
CalcHEP/micrOMEGAs will use the downloaded values for particle widths instead of
its  automatic  calculation. Below is an example of a SLHA file record and of the SLHAplus
instructions to read them 

{\small
\begin{verbatim}
         File records                        SLHAPlus reading 
                               
BLOCK STOPmix # stop mixing matrix       
1 1 5.37975095e-01 # O_{11}          Zt11=slhaVal("STOPmix",MZ,2,1,1);
1 2 8.42960733e-01 # O_{12}          Zt12=slhaVal("STOPmix",MZ,2,1,2);
2 1 8.42960733e-01 # O_{21}          Zt21=slhaVal("STOPmix",MZ,2,2,1);
2 2 -5.37975095e-01 # O_{22}         Zt22=slhaVal("STOPmix",MZ,2,2,2);
\end{verbatim}
}
Here the first argument of {\tt slhaVal} is the name of {\it block}, the second one is the  {\it scale}
parameter (in this example the scale parameter is not specified in BLOCK), the third parameter fixes the number
of {\it key} parameters,   the key parameters themselves follow.  
 
Another important option of the {\it SLHAplus} package  
is a tool for diagonalizing mass matrices which occur in models
with complicated  multiplet structure.     

\subsection{ Generation of model files by LanHEP.}
The  {\tt LanHEP}\cite{Semenov:2008jy} package allows  automatic generation of CalcHEP
model files. It starts from model definition in terms of particle multiplets
and performs substitution of physical particle fields  in multiplets.
LanHEP also checks at the symbolic level the   absence of linear terms  and at the
numerical level the  absence  of off-diagonal terms in the quadratic part of Lagrangian.
Also LanHEP compares diagonal quadratic terms with declared particle
masses.   Using LanHEP  allows to avoid a large number of mistakes
which could appear in the derivation of Feynman rules by hand. 

We present here part of a LanHEP input file for the IDM model describing only
new particles and new interactions beyond  the Standard Model.
\begin{verbatim} 
parameter MHX=111,MH3=222,MHC=333.  % Declaration of new masses  
parameter laL=0.01, la2=0.01.       % Declaration of new couplings 

%mu^2 as a function of masses
parameter mu2=MHX**2-laL*(2*MW/EE*SW)**2.   
% constraints for couplings
parameter la3=2*(MHC**2-mu2)/(2*MW/EE*SW)**2.   
parameter la5=(MHX**2-MH3**2)/(2*MW/EE*SW)**2.
parameter la4=2*laL-la3-la5.                   
% Declaration of new particles 
scalar '~H3'/'~H3':('odd Higgs',pdg 36, mass MH3, width wH3 = auto).
scalar '~H+'/'~H-':('Charged Higgs',pdg 37,mass MHC,width wHC=auto).
scalar '~X'/'~X':('second Higgs',pdg 35,mass MHX,width wHX=auto).
% second Higgs doublet 
let h2 = { -i*'~H+',  ('~X'+i*'~H3')/Sqrt2 },
    H2 = {  i*'~H-',  ('~X'-i*'~H3')/Sqrt2 }.
% covariant derivative. 
% Here B1 -U(1)  gauge field, 
% WW={W-,W3,W+} SU(2) triplet 
let Dh2^mu^a = (deriv^mu+i*g1/2*B1^mu)*h2^a +
         i*g/2*taupm^a^b^c*WW^mu^c*h2^b.
let DH2^mu^a = (deriv^mu-i*g1/2*B1^mu)*H2^a
        -i*g/2*taupm^a^b^c*{'W-'^mu,W3^mu,'W+'^mu}^c*H2^b.

lterm DH2*Dh2.                 % Kinematic and other terms.  
% Below  h1 is standard Higgs doublet.  
lterm -mu2*h2*H2.
lterm -la2*(h2*H2)**2.
lterm -la3*(h1*H1)*(h2*H2).
lterm -la4*(h1*H2)*(H1*h2).
lterm -la5/2*(h1*H2)**2 + AddHermConj.
\end{verbatim}
The complete LanHEP input file for the IDM can be found  in the directory 
\verb|IDM/lanhep|. Indeed all models treated by micrOMEGAs were constructed
with LanHEP.  Compilation of models by LanHEP can be done by the command \\
\verb|     lhep  <source file>  -ca -evl 2|

\bibliographystyle{ws-procs9x6}

\end{document}